\documentclass[twocolumn]{aastex63}
\usepackage{xcolor}
\usepackage{hyperref}

\usepackage [caption=false]{subfig}
\usepackage{natbib}
\usepackage{url}

\newcommand\Nu{\textit{NuSTAR}}
\newcommand\xmm{\textit {XMM-{Newton}}}

\begin{document}

\title{Sudden emergence of a low-frequency hard X-ray lag in the Seyfert 1 galaxy Mrk 1044}
\correspondingauthor{Jia-Lai Kang \& Jun-Xian Wang} \email{ericofk@mail.ustc.edu.cn, jxw@ustc.edu.cn}

\author{Jia-Lai Kang} 
\affiliation{Department of Astronomy, University of Science and Technology of
China, Hefei, Anhui 230026, China}
\affiliation{School of Astronomy and Space Science, University of Science and Technology of China, Hefei 230026, China}
\author{Jun-Xian Wang}
\affiliation{Department of Astronomy, University of Science and Technology of
China, Hefei, Anhui 230026, China}
\affiliation{School of Astronomy and Space Science, University of Science and Technology of China, Hefei 230026, China}
\affiliation{College of Physics, Guizhou University, Guiyang, Guizhou 550025, People’s Republic of China}

\begin{abstract}
Hard X-ray lags, where low frequency variations in the hard X-ray band lag behind those in the soft band, have been detected in many active galactic nuclei (AGNs) and are generally attributed to the inward propagation of accretion-flow fluctuations through an extended corona. In a long \xmm~observation of the Seyfert 1 galaxy Mrk 1044, we found a remarkable transition in the lag behavior within a single exposure, while the X-ray flux and spectral shape remained largely unchanged. During the first 60 ks, no significant hard X-ray lag was detected, whereas in the subsequent 60 ks, a pronounced lag emerged. The lag was so prominent that a large-amplitude flux variation event during the lag-detected interval, characterized by a gradual dimming followed by recovery, produced a remarkable clockwise loop in the flux-softness diagram. The sudden appearance of the hard X-ray lag suggests that the X-ray corona underwent a rapid transition from a compact to an extended configuration. This scenario is further supported by two independent observational signatures: (1) the variability became noticeably smoother, with a redder power spectral density (PSD), during the lag-detected interval, and (2) the broad Fe K$\alpha$ line profile became narrower and the reflection continuum weaker. These findings highlight the diagnostic power of tracking rapid changes in hard X-ray lags for probing the physical structure and evolution of AGN coronae, and demonstrate that identifying prominent loops in the flux-softness diagram provides an effective way to locate intervals with significant hard X-ray lags.

\end{abstract}

\keywords{Galaxies: active – Galaxies: nuclei  – X-rays: galaxies }

\section{Introduction} \label{sec:intro}
Active galactic nuclei (AGNs) emit powerful X-rays from a hot and compact region near the central supermassive black hole, the so-called X-ray corona \citep{Haardt_1991, Haardt_1993}. Energetic electrons in the corona upscatter ultraviolet/optical seed photons from the accretion disk into the X-ray band, producing the observed power-law–shaped X-ray continuum with a high-energy cutoff \citep{Nandra_1994, Zdziarski_2000, Kang_2022}. In addition to the primary continuum, prominent reflection features are often present in the spectra, including both broad and narrow Fe K$\alpha$ lines around 6.4 keV \citep[e.g.,][]{Pounds_1990, Tanaka_1995, Wang_1999} and a Compton hump peaking near 30 keV \citep[e.g.,][]{George_1991, Ricci_2011, Walton_2014, Panagiotou_2020}. Moreover, an excess of emission over the extrapolated power-law continuum below 2 keV—the so-called soft X-ray excess—has been detected in a large fraction of AGNs \citep{Gierlinski_2004, Bianchi_2009}. The origin of the soft excess remains under debate \citep{Boissay_2016, Garcia_2019, Zoghbi_2023}. Proposed explanations include emission from an additional warm corona \citep{Czerny_2003, Done_2012, Petrucci_2020}, relativistic ionized reflection \citep{Crummy_2006, Walton_2013, Jiang_2019}, or a combination of both \citep{Chen_2025,Chen_2025b}.

\par Although the X-ray corona has been studied for decades, its physical nature remains poorly understood. Temporal variability, especially time lags between different X-ray energy bands, provides a powerful diagnostic of the geometry and physical processes of the corona. The X-ray spectral components often vary in a correlated or coherent manner across energies \citep[e.g.][]{Epitropakis_2017, Ren_2025}, implying causal connections between different emission regions. In particular, frequency-dependent time lags have been widely observed and can be broadly categorized into two types: high-frequency soft lags and low-frequency hard lags. Over the past decades, studies have revealed that variations in softer bands lag behind those in harder bands at frequencies higher than $\sim10^{-4} \rm \, Hz$ \citep[e.g.,][]{Fabian_2009, Zoghbi_2010, Emmanoulopoulos_2011, DeMarco_2013, Caballero_2018}. These so-called soft lags are primarily attributed to X-ray reverberation \citep{Wilkins_2013, Uttley_2014}, as further supported by the detection of similar time lags between the Fe K$\alpha$ line and the continuum \citep[e.g.,][]{Zoghbi_2012, Kara_2013, Marinucci_2014} and between the reflection hump and the continuum \citep[e.g.,][]{Zoghbi_2014, Kara_2015, Zoghbi_2021}. In contrast, hard lags, where the harder band lags behind the softer band, have been detected at lower frequencies in dozens of AGNs \citep{Papadakis_2001, McHardy_2004, Arevalo_2008, Alston_2014, Kara_2016, Lobban_2018}, and are likely caused by the inward propagation of accretion-flow fluctuations through an extended corona \citep{Kotov_2001, Arevalo_2006}.

Remarkably, investigations of the time- or flux-dependent evolution of high-frequency lags have revealed the highly dynamic nature of the corona \citep[e.g.,][]{Alston_2020, Hancock_2022, Wilkins_2023, Nakhonthong_2024, Yu_2025}.
In contrast, studies focusing on the variability of low-frequency lags remain scarce \citep{Alston_2013, Kara_2013b}, and the temporal evolution of these slow-varying signals is still poorly characterized.

Mrk 1044 is a low-redshift narrow-line Seyfert 1 galaxy ($z=0.017$; \citealt{Koss_2022}) with a black hole mass of $M_{\rm BH} \sim 3 \times 10^{6} M_{\odot}$ \citep{Du_2015}. Previous studies have investigated its X-ray properties, including the soft excess \citep{Dewangan_2007}, ultrafast outflows \citep{Krongold_2021, Xu_2023}, X-ray/UV correlations \citep{Barua_2023}, and high-frequency soft and low-frequency hard time lags \citep[from earlier 2016 observations,][]{Mallick_2018}.

In this paper, we report the sudden emergence of a hard X-ray lag in Mrk 1044, based on comprehensive timing analysis of high-quality joint \xmm~and \Nu~observations in 2018. During the 2018 observation, we identified a remarkable loop in the time-resolved ``spectral-softness-ratio -- count-rate'' diagram. Detailed analysis shows that this loop is caused by a prominent low-frequency hard X-ray lag, which was absent during the first 60 ks of the observation but emerges in the subsequent interval.

\begin{figure*}
\centering
\subfloat{\includegraphics[width=0.95\textwidth]{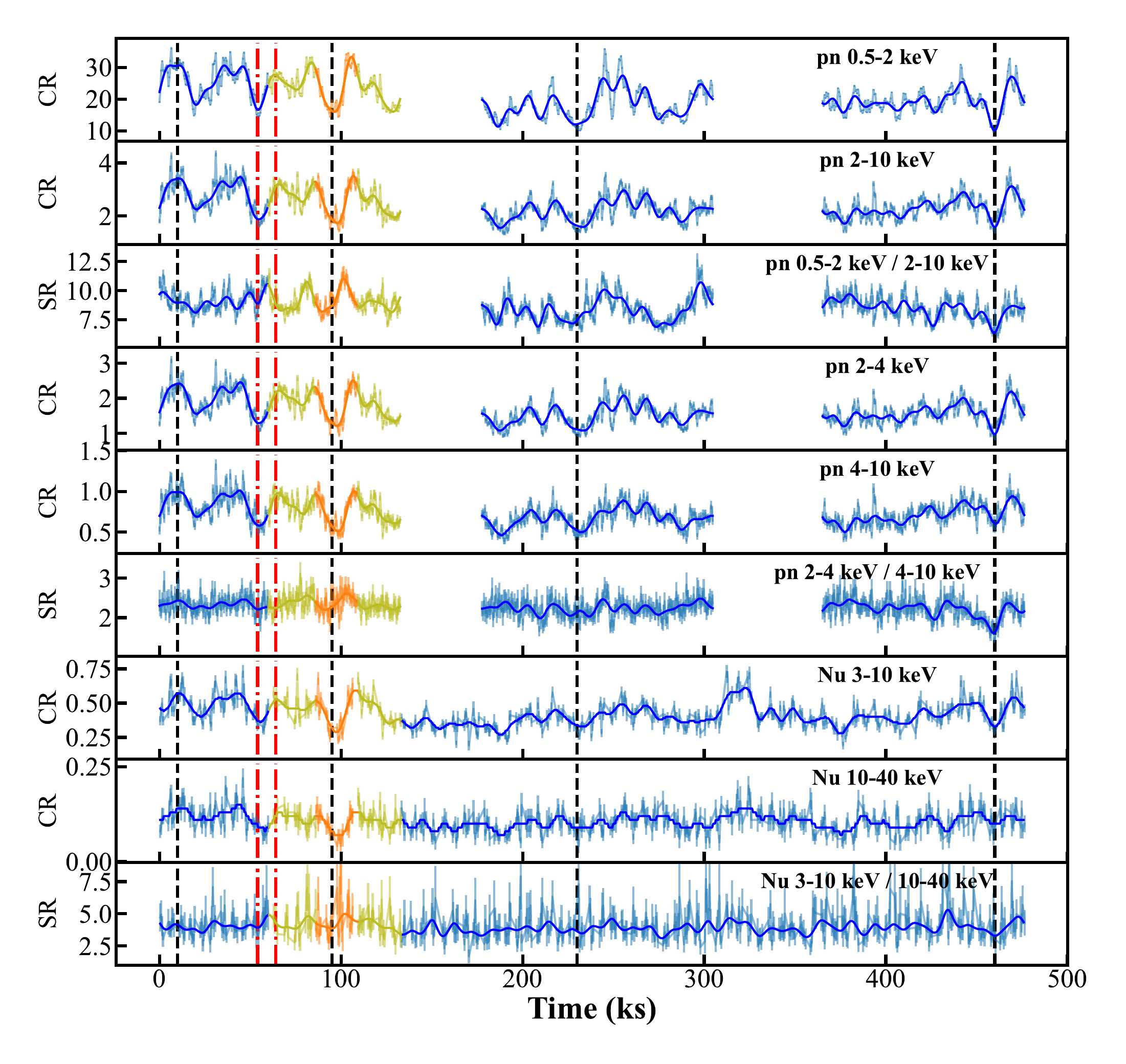} }
\caption{\label{fig:LC}Net count rate (CR, in units of counts s$^{-1}$) and softness ratio (SR) curves in different energy bands for the three \xmm~and one \Nu~observations. Light-colored points represent the original data, while solid lines show the low-frequency ($\le 10^{-4}$~Hz) variations after filtering out high-frequency fluctuations. The first \xmm~observation is divided into two halves: the first half (blue) and the latter half (orange and olive), each spanning $\sim 60$~ks. The four vertical black dashed lines indicate notable peaks and troughs in the CR curves across different periods, while the two vertical red dashdotted lines show the possible timing when the lag start to appear (see \S \ref{S:timing}). The hard X-ray lag is most pronounced during the orange interval, producing a prominent loop in the time-resolved flux–softness diagram (Fig.~\ref{fig:CRSR}). }
\end{figure*}

\section{Data Reduction}

\par A deep joint \xmm~\citep{Jansen_2001} and \Nu~\citep{Harrison_2013} campaign of Mrk 1044 was conducted in 2018, including three \xmm~observations (ID: 0824080301, 0824080401 and 0824080501; $\sim 140$ ks duration each), and one \Nu~observation (ID=60401005002, $\sim 560$ ks duration). 

\par For the \xmm~observations, we focus on the EPIC-pn data \citep{Struder_2001}, all taken in Small Window mode with the Medium filter. The raw data were reduced using the \xmm~Science Analysis System (SAS, version 20.0.0) with the Current Calibration Files (CCF). No intervals were affected by background flares based on the criterion of \citet{Kang_2024}, allowing us to derive consecutive, evenly sampled light curves for each observation. Source events were extracted from a circular region of radius 60\arcsec, with nearby source-free regions used for background. The mean 0.5–10 keV count rate is 27 cts/s, and pile-up is negligible according to the SAS task \textsc{epatplot}. Light curves were extracted using \textsc{evselect} with a 500 s bin and corrected with \textsc{epiclccorr} for background and instrumental effects. Spectra and response files were obtained with \textsc{especget} and grouped to a minimum of 50 counts per bin using \textsc{specgroup}. Although a cross-calibration issue exists between EPIC-pn and \Nu, which can be empirically corrected in the 3–10 keV band using \textsc{arfgen} \footnote{\url{https://xmmweb.esac.esa.int/docs/documents/CAL-TN-0230-1-3.pdf}}, we do not apply this correction here, as we focus on the 0.5–10 keV EPIC-pn data.

\par \Nu~data were reduced with \textsc{nupipeline} (HEASoft 6.32.1) and calibration files version 20210824. Source and background events were extracted from circular regions of 60\arcsec\ radius. Light curves from FPMA and FPMB were combined and background-subtracted using \textsc{lcmath}, with a 500 s bin. To ensure quasi-simultaneity, \Nu~spectra were extracted in the overlapping Good Time Intervals (GTIs) with each \xmm~observation. Spectra were grouped to a minimum of 50 counts per bin using \textsc{grppha}. Using a finer binning, e.g., a minimum of 20 counts per bin, can significantly bias spectral fitting of the NuSTAR data, likely due to the high background fraction at hard X-ray energies \citep[see Fig. 2 in][]{Liao_2024}. Given the steep X-ray spectrum of this high Eddington ratio NLS1 ($\Gamma \sim 2.5$) and the dominance of background above 40 keV, only the 3–40 keV \Nu~data are used.

\par Since there are $\sim30$ ks gaps between the three \xmm~observations (Fig. \ref{fig:LC}), frequency-domain analyses were performed individually for each observation with consecutive and evenly sampled light curves. For the \Nu~light curves, which contain multiple gaps due to Earth occultation, we employed a Gaussian process method to generate continuous light curves using the \textsc{pylag} code \citep{Wilkins_2019}, using the ``RationalQuadratic'' kernel function\footnote{\url{https://scikit-learn.org/stable/modules/generated/sklearn.gaussian_process.kernels.RationalQuadratic.html}}. We note that these estimated continuous light curves are only used for visualization in Fig. \ref{fig:LC} and \ref{fig:CRSR}, not used for timing analysis and lag estimation.

\section{Results}

\subsection{Light curves and flux-softness diagrams}

\begin{figure*}
\centering
\subfloat{\includegraphics[width=0.95\textwidth]{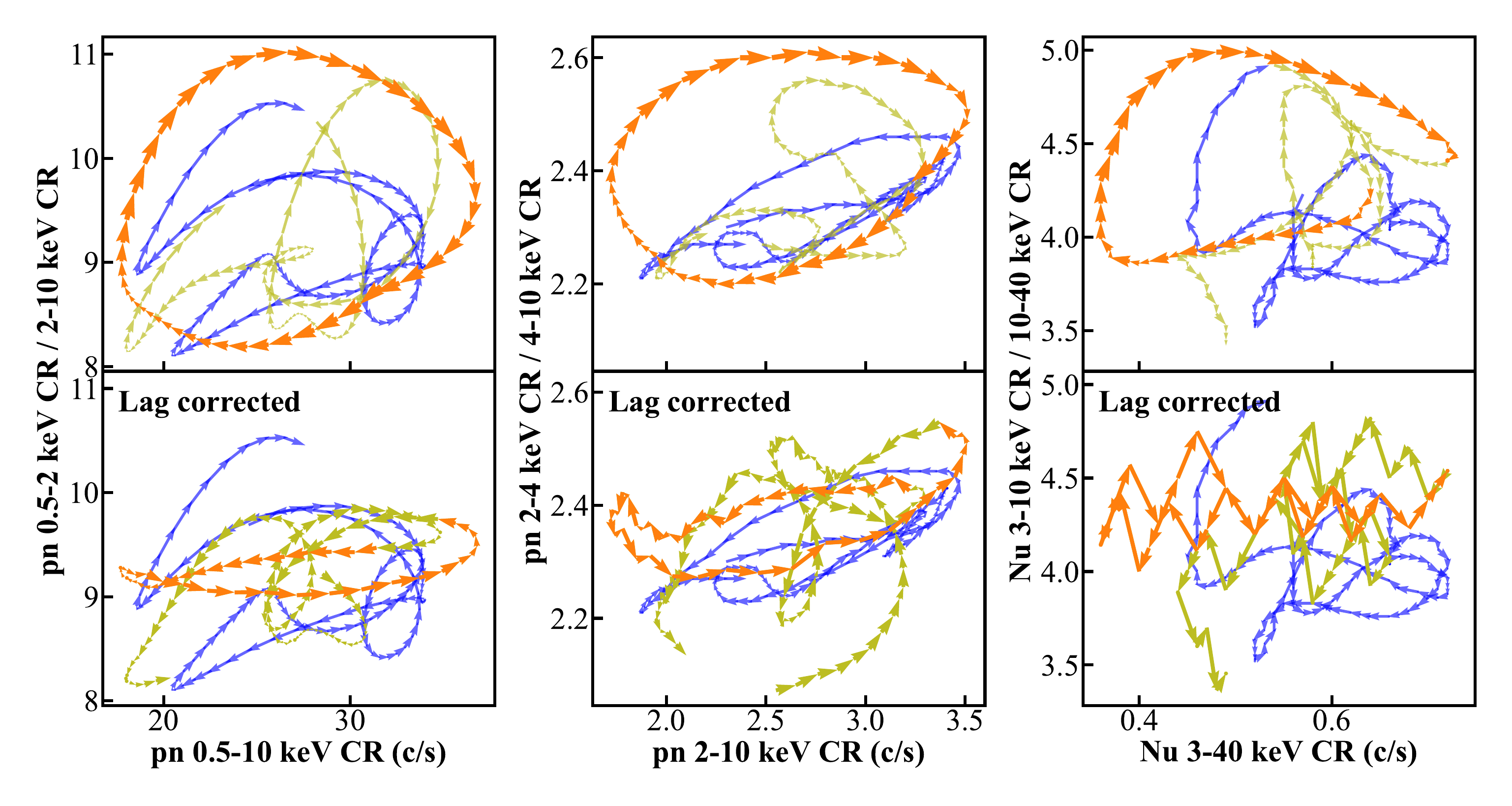}
 }
\caption{\label{fig:CRSR}Low-frequency ($\le 10^{-4}$~Hz) CR–SR diagrams for the first XMM–Newton observation (before $\sim$130~ks in Fig.~\ref{fig:LC}), with arrows indicating the time sequence. The top panels show the observed CR–SR relations in different energy bands: from left to right, 0.5–2~keV CR / 2–10~keV CR versus 0.5–10~keV CR, 2–4~keV CR / 4–10~keV CR versus 2–10~keV CR, and 3–10~keV CR / 10–40~keV CR versus 3–40~keV CR. The curves are color-coded as in Fig.~\ref{fig:LC}. During the period marked in orange, prominent clockwise loops appear in all three diagrams, indicating strong hard X-ray lags between the corresponding bands. The bottom panels show the CR–SR diagrams after correcting for the measured hard lags (1.5, 1.0, and 1.0~ks for each band pair, respectively), where the loops collapse, validating the lag measurements.
}
\end{figure*}

\par We present in Fig. \ref{fig:LC} the count-rate (CR) light curves in different energy bands, together with the corresponding softness ratio ($\mathrm{SR} = \rm CR_{\mathrm{soft}} / CR_{\mathrm{hard}}$) curves. We use the softness ratio rather than the more commonly used hardness ratio, so that the SR curves display peaks and troughs consistent with those in the count-rate curves, as a consequence of the ``softer-when-brighter'' behaviour \citep{Sobolewska_2009, Wu_2020}. To highlight the low-frequency variability, we overplot the low-frequency ($\leq 10^{-4}$ Hz) components (solid lines) obtained after filtering out the high-frequency fluctuations\footnote{This is achieved by performing a Discrete Fourier Transform (DFT), setting the coefficients of frequency bins above $10^{-4}$ Hz to zero, and then applying an inverse DFT. }. 

\par It is evident that during the latter half of the first \xmm~observation (highlighted in orange and olive), the count-rate curves in the harder bands systematically and significantly lag behind those in the softer bands, indicating a pronounced hard X-ray lag. Consequently, the softness-ratio curves lead the corresponding count-rate curves. We further mark four characteristic extrema (peaks or troughs) with vertical dashed lines. By visual inspection, the hard lag is absent during the first half of the first \xmm~observation, suddenly appears in the latter half, weakens but persists in the second \xmm~observation, and disappears toward the end of the entire monitoring campaign.

\par This behavior could also be clearly illustrated in the time-resolved flux–softness diagram. In Fig.~\ref{fig:CRSR}, we plot the low-frequency SR versus CR for the first \xmm~observation, with arrows indicating the time sequence. A prominent clockwise convoluted loop is evident during the latter half of the observation (color-coded in orange). Such a loop signifies the presence of a low-frequency lag \citep[e.g.,][]{Weng_2021, Wilkins_2023}. Here, the clockwise sense of the loop is interpreted as SR variations leading those in the broad-band CR (i.e., the harder band lags behind the softer band), whereas a counterclockwise loop would indicate a soft lag. In contrast, during the first half of the observation (shown in blue), no significant loop is seen, suggesting the absence of a measurable lag.

\par Furthermore, Fig.~\ref{fig:CRSR} also reveals similar loops across different energy bands, indicating that the hard X-ray lag arises in the broad-band continuum rather than solely between specific spectral components (e.g., the power-law continuum and the soft excess). In addition, the mean CR and SR values during the loop are nearly identical to those of the first part (shown in blue), suggesting that the time-averaged flux and spectral shape remain largely unchanged despite the sudden emergence of the hard lag.

It is worth noting that this peculiar event was first identified serendipitously when we examined the time-resolved flux–softness diagrams of X-ray–bright AGNs in search of possible eclipse events \citep[e.g.,][]{Kang_2023}, following the approach we proposed in \citet[][see their Fig.~17]{Zhou_2025SCPMA}. This incidental discovery demonstrates that inspecting time-resolved flux–softness diagrams for loop-like structures can serve as an effective means to identify pronounced hard X-ray lags and their potential variability. A systematic search for such events in a large AGN sample will be carried out in future work.

\subsection{Time lags} \label{sec:lags}

\par
\begin{figure}
\centering
\subfloat{\includegraphics[width=0.45\textwidth]{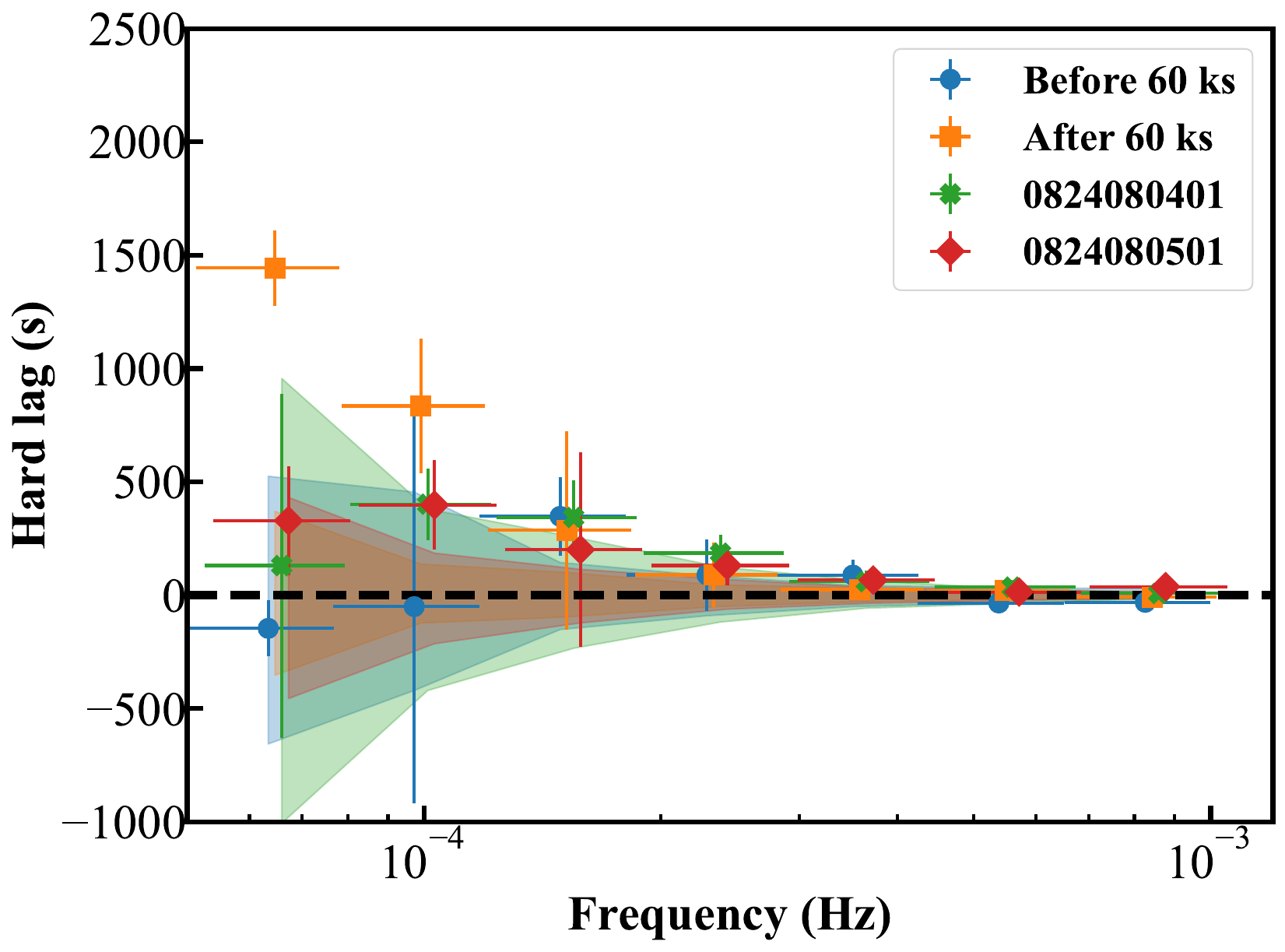}}\\
\subfloat{\includegraphics[width=0.45\textwidth]{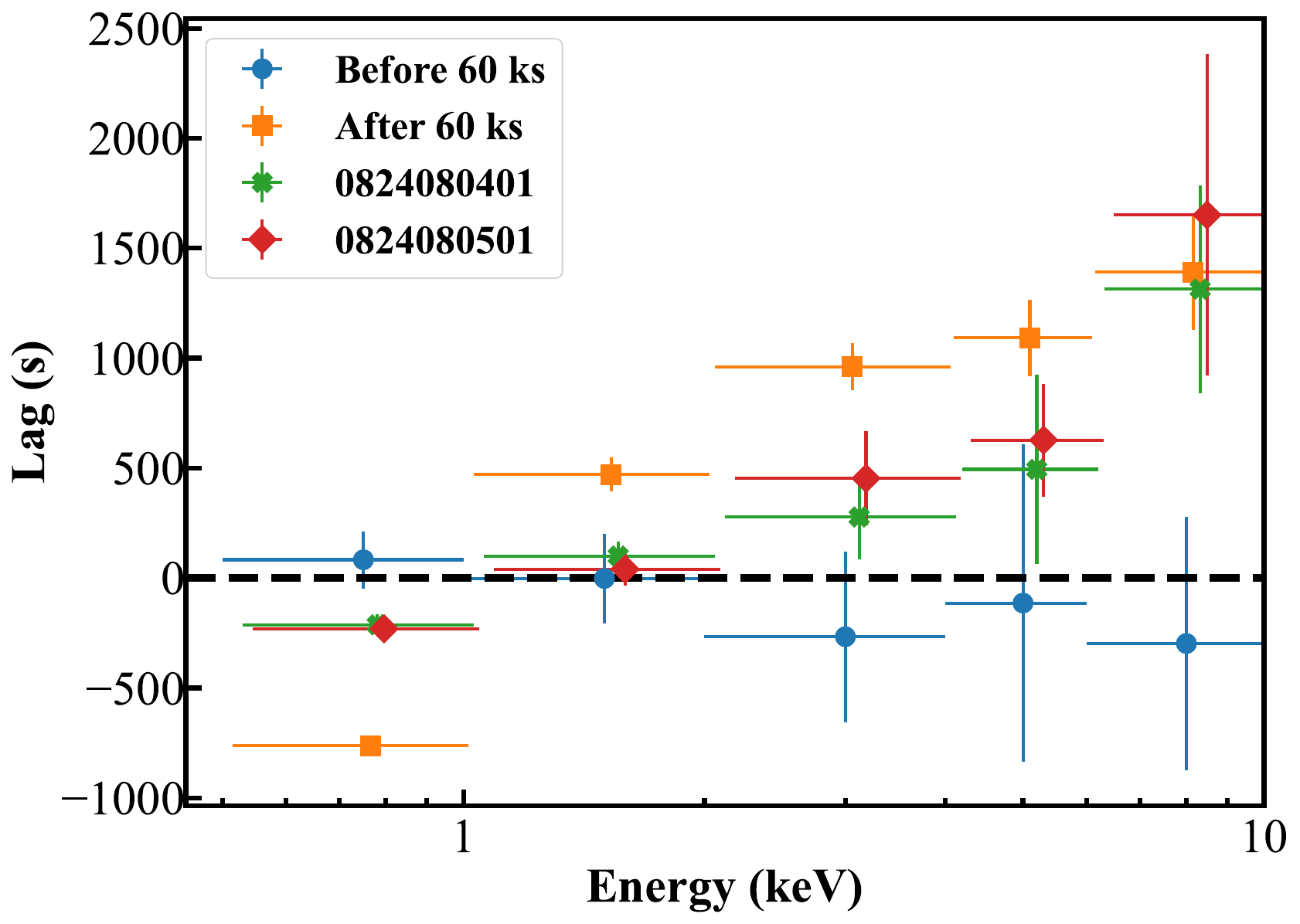}}
\caption{\label{fig:lags}Top panel: hard X-ray time lags (2–10~keV relative to 0.5–2~keV, EPIC-pn) as a function of temporal frequency, measured before and after 60~ks in observation 0824080301 (blue circles and orange squares), and during observations 0824080401 (green crosses) and 0824080501 (red diamonds). The error bars show the 1$\sigma$ uncertainties derived from \textsc{pylag} based on the coherence in each frequency bin. The shaded areas show the 1$\sigma$ null distribution of the measured lags from Monte Carlo simulations under the null hypothesis of zero intrinsic lag, constructed to match the measured power spectral density and coherence spectrum of each band (see \S \ref{sec:lags}); the lags measured after 60~ks lie outside this null distribution at $>99\%$ confidence in the three lowest frequency bins.
Bottom panel: lag–energy spectra calculated within the low-frequency range ($5.0\times10^{-5}$–$1.0\times10^{-4}$~Hz) using the 0.5–10~keV band as the reference. The X-values of the data points are slightly offset for clarity.
}
\end{figure}

 We then directly measure the frequency- and energy-dependent time lags using \textsc{pylag}, with the methodology summarized in \citet[][Section 2.1]{Uttley_2014} \footnote{\citet{Epitropakis_2016, Epitropakis_2017} proposed an alternative method to estimate the Fourier-based time lags, by splitting the observed light curves into shorter data segments and averaging the resulting cross-periodograms \citep{Priestley_1981}, which could reduce the measurement biases. However, the observations in this work are not long enough for applying this method.}. Note that no Gaussian process is involved here in calculating the lags. We divide the first XMM-Newton observation into two segments, ``before 60 ks'' and ``after 60 ks'', and compute the time lags in each separately. The dividing point is approximately the midpoint of the observation, and coincides with the apparent onset of a significant lag (located between the two red vertical dash-dotted lines in Fig. \ref{fig:LC}). See \S \ref{S:timing} for a more detailed discussion. The top panel of Fig.~\ref{fig:lags} shows the frequency-resolved time lags between the EPIC-pn 2–10~keV and 0.5–2~keV count-rate curves (with a time bin of 500~s), computed in seven logarithmically spaced frequency bins between $5.0\times10^{-5}$ and $1.0\times10^{-3}$~Hz. This analysis again confirms the scenario in which a low-frequency hard X-ray lag suddenly emerges after $\sim$60~ks in the first \xmm~observation, and then weakens during the subsequent two observations.

\par To further quantify the significance of the detection and non-detection of the time lags, we perform a null-hypothesis Monte Carlo test following the general approach of \citet{Zoghbi_2010} and \citet{Mallick_2021}. For each observation period, we first fit the power spectral density (PSD) of both the 0.5--2~keV and 2--10~keV EPIC-pn light curves with a broken power law, and measure the coherence spectrum $\gamma^2(f)$ between the two bands \citep[e.g.,][]{Uttley_2014}. We then generate a single long simulated light curve pair of $10^{3}$ times the observed length, following the method of \citet{Timmer_1995}: the soft-band Fourier transform is drawn from the best-fit soft-band PSD, and the hard-band Fourier transform is constructed at each frequency as the sum of a coherent component (a fraction $\gamma^2(f)$ of the power, phase-shifted from the soft band by zero lag, i.e., the null hypothesis of no intrinsic lag) and an independent component (a fraction $1-\gamma^2(f)$, drawn from its own red-noise realization following the best-fit hard-band PSD). This construction reproduces both the measured PSD of each band and the measured coherence spectrum, while enforcing zero intrinsic lag between the bands. Poisson noise consistent with the observed count rates and fractional rms is added to both light curves. We then randomly draw 1000 sub-segments of the same length as the observed light curve from this long pair, and measure the lag spectrum of each sub-segment using the same cross-spectrum method applied to the data. The resulting distribution in each frequency bin is the null distribution of the measured lag, which is shown as the shaded regions in Fig~\ref{fig:lags}. For the ``after 60~ks'' interval, the observed lags lie outside this null distribution at $>99\%$ confidence ($\gtrsim 2.8\sigma$) in the three lowest frequency bins. The lags measured in the ``before 60~ks'' interval are consistent with the null hypothesis at all frequencies.

\par The bottom panel of Fig.~\ref{fig:lags} presents the lag–energy spectra, calculated within the low-frequency range of $5.0\times10^{-5}$–$1.0\times10^{-4}$~Hz. For each energy band, the lag was computed relative to the broad 0.5–10~keV reference band \citep{Uttley_2014}. Consistent with Fig.~\ref{fig:CRSR}, the lag–energy spectra reveal that the hard X-ray lag spans a wide energy range during this special epoch. In particular, significant hard lags are detected even within the power-law–dominated bands, suggesting that this event cannot be fully explained by uncorrelated variability between the soft excess and the primary powerlaw continuum.

\par From the lag–energy spectra, we further estimate typical low-frequency hard lags between representative energy bands. The hard lags are approximately 1.5, 1.0, and 1.0~ks between EPIC-pn 2–10~keV and 0.5–2~keV, 4–10~keV and 2–4~keV, and \Nu~10–40~keV and 3–10~keV, respectively. We then corrected the light curves for these lags by offsetting them by 3, 2, and 2 time bins, recalculated the SR curves, and replotted them in the bottom panels of Fig.~\ref{fig:CRSR}. The disappearance of the giant loops after this correction provides a validation of the measured time lags \citep[see also Fig.~6 in, e.g.,][]{Weng_2021}.

\subsection{Timing of the Transition} \label{S:timing}
\par In the preceding analysis, we divided the first \xmm~observation into two intervals, before and after 60~ks, assuming that the sudden transition from no measurable hard lag to a pronounced low-frequency hard lag occurs around 60~ks. In this subsection, we present a more detailed analysis to support this assumption.

\par
\begin{figure}
\subfloat{\includegraphics[width=0.45\textwidth]{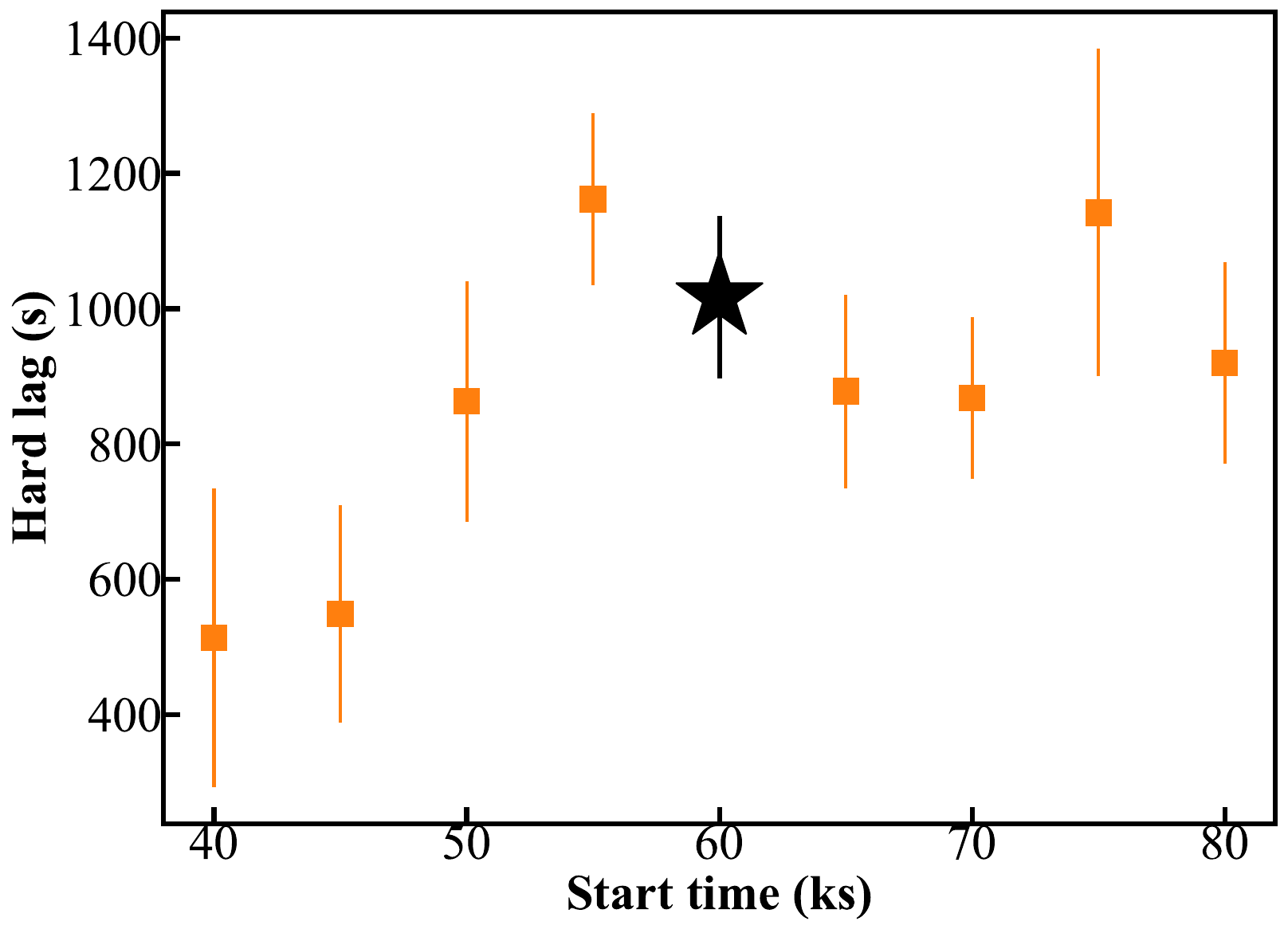}}
\caption{\label{fig:start_of_event}Lags between the 2–10~keV and 0.5–2~keV light curves in the frequency range [$5\times10^{-5}$, $1\times10^{-4}$]~Hz, calculated for the latter part of observation 0824080301 with different start times. Each data point represents the cumulative time lag computed over the segment of the light curve from the given start time to the end of observation 0824080301. For example, the data point at a start time of 40 ks is derived from the light curve spanning 40 ks to $\sim$130 ks in Fig. \ref{fig:LC}. Note that the error bars are not mutually independent, as adjacent data points are computed from overlapping segments of the light curve.
In this work, we adopt 60~ks (marked by the black star) to divide the observation into two segments with comparable exposure times.
}
\end{figure}

 In Fig.~\ref{fig:LC}, we add two red vertical dash-dotted lines marking a trough (at 54~ks) and a subsequent peak (at 64~ks) near the presumed transition point (around 60~ks). No significant lag is observed between the CR curves at the trough, whereas a small lag becomes apparent at the peak, suggesting that the onset of the lag likely occurred between 54 and 64~ks. To quantify this further, we calculated the low-frequency ($5\times10^{-5}$–$1\times10^{-4}$~Hz) lag between the 2–10~keV and 0.5–2~keV light curves in the latter part of the first \xmm~observation, using a series of different start times. As shown in Fig.~\ref{fig:start_of_event}, the measured hard lags for the intervals starting at 40 and 45~ks are considerably smaller, as these segments still include intervals without or with very small lags. In contrast, the lag stabilizes at interval start time $\geq$55~ks. Although the limited data quality prevents us from pinpointing the exact onset, the transition appears to have occurred rapidly between 50 and 60~ks. We therefore adopt 60~ks to divide the first \xmm~observation into two intervals of comparable exposure for a fair comparison.

\subsection{What Drives This Special Event?}

Compared with the previously reported events of varying hard X-ray lags \citep{Alston_2013, Kara_2013b}, the special event in Mrk~1044 is strikingly distinct, exhibiting a rapid variation of the lag from $\sim$0~ks to 1.5~ks on a short time-scale of only a few kiloseconds\footnote{The $M_{\rm BH}$ of Mrk~1044 is $3 \times 10^{6} M_{\sun}$ \citep{Du_2015}, corresponding to a Schwarzschild radius of $R_{s} \sim 9 \times 10^{6}$~km, so 1.5~ks corresponds to $\sim 50 R_{s}$ assuming light-speed propagation.}, while the average flux remains nearly constant. To investigate the physical origin of this special period, we further compare the spectra at different epochs.

\par Previous studies of Mrk~1044 have shown that its X-ray spectrum is rather complex, exhibiting both relativistic reflection \citep{Mallick_2018} and ionized absorption features associated with ultrafast outflows \citep{Krongold_2021, Xu_2023}. Since our focus here is on spectral shape variations across different epochs, we adopt the spectral ratio method to directly illustrate these changes \citep{Zhangjx2018, Kang_2021}. In brief, we fit each spectrum with a simple model, $tbabs \times (bbody + powerlaw)$, using \textsc{xspec} \citep{Arnaud_1996}, obtain the unfolded spectra, i.e., the spectra with instrumental response deconvolved (\textit{ufspec} in \textsc{xspec}), and compute the ratios of the unfolded spectra at each epoch relative to that at the special epoch exhibiting the pronounced hard X-ray lag. We note that, although the unfolded spectra are model-dependent, the spectral ratio method is largely insensitive to the choice of continuum model and provides a reliable way to reveal variations in the broad-band continuum properties \citep{Zhangjx2018}.

\begin{figure}
\centering
\subfloat{\includegraphics[width=0.45\textwidth]{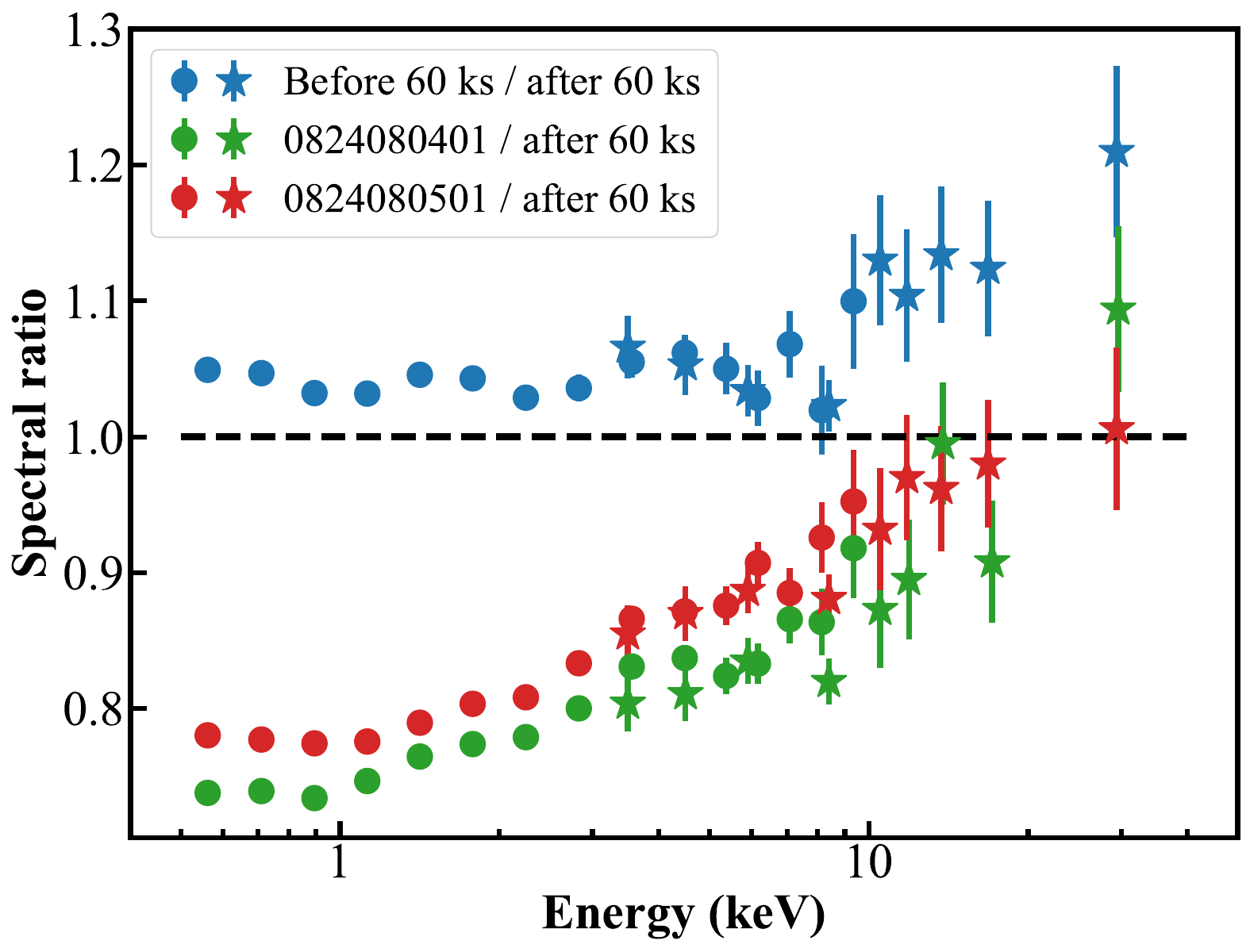}}
\caption{\label{fig:spectral_ratio} Ratios of the unfolded spectra (with instrumental response deconvolved) at different epochs to that of the special epoch showing a strong hard X-ray lag (after 60~ks in observation 0824080301). No model is involved in the ratios directly; a simple model \textit{tbabs$\times$(bbody+powerlaw)}, with all parameters free, is used solely to obtain the unfolded spectra via \textit{ufspec} in \textsc{xspec}. Following \citet{Zhangjx2018}, the unfolded spectrum is treated as an approximation of the intrinsic broad-band spectrum when only the broad-band continuum shape is of interest, so the ratio between epochs directly traces variations in the continuum. For example, the green points show that the spectrum in observation 0824080401 is fainter (ratio $\lesssim$ 1.0 in nearly all bands) and harder (i.e., the ratio increases toward higher energies) compared with the reference epoch. Circles and stars denote the XMM–Newton EPIC-pn and NuSTAR data, respectively.
}
\end{figure}

\par The spectral ratios are shown in Fig.~\ref{fig:spectral_ratio}. Compared with the reference period exhibiting the strongest hard X-ray lag, the spectrum during the first 60~ks of 0824080301 is only $\sim 5\%$ brighter, and the spectral shape is almost identical. The \Nu~data show a slight excess above $\sim 10$~keV, which could be attributed to a stronger reflection component (although no corresponding signal is seen below 2~keV where the ionized reflection component would appear) or a higher coronal temperature.
The near-constant flux and spectral shape effectively rule out an eclipsing scenario \citep[e.g.,][]{Gallo_2021, Kang_2023}, in which an absorber successively obscures different emitting regions, producing a time lag.

Meanwhile, the spectra in the later two observations are significantly fainter and harder, consistent with the well-known “softer-when-brighter” behavior observed in Seyfert galaxies \citep{Markowitz_2003, Sobolewska_2009, Wu_2020}. At energies above $\sim 1$~keV, the spectral ratios are consistent with straight lines in log space, indicating variations in the photon index of the power-law continuum. Below $\sim 1$~keV, the ratios flatten, likely because the soft excess does not respond in concert with the power-law continuum \citep[e.g.,][]{Ren_2025}.

\par Hard X-ray lags are generally interpreted as arising from the inward propagation of fluctuations through an extended corona, with the “hardest” X-ray emission originating from the innermost regions \citep{Kotov_2001, Arevalo_2006}. Within this framework, the observed variation of the hard X-ray lag implies that either the coronal size, the propagation speed of the fluctuations, or both are changing rapidly. 
While the propagation speed cannot be directly constrained, independent measurements of the coronal size in each period, e.g., from reflection modelling or high-frequency reverberation time lags \citep[e.g.,][]{Ursini_2020b, Alston_2020, Wilkins_2023, Yu_2025}, could be helpful to break the degeneracy between changes in coronal geometry and in the propagation speed. In the following, we focus primarily on possible changes in the coronal geometry, while also discussing the extent to which the present data can provide independent constraints on changes in the coronal size between different periods.

\begin{figure}
\centering
\subfloat{\includegraphics[width=0.45\textwidth]{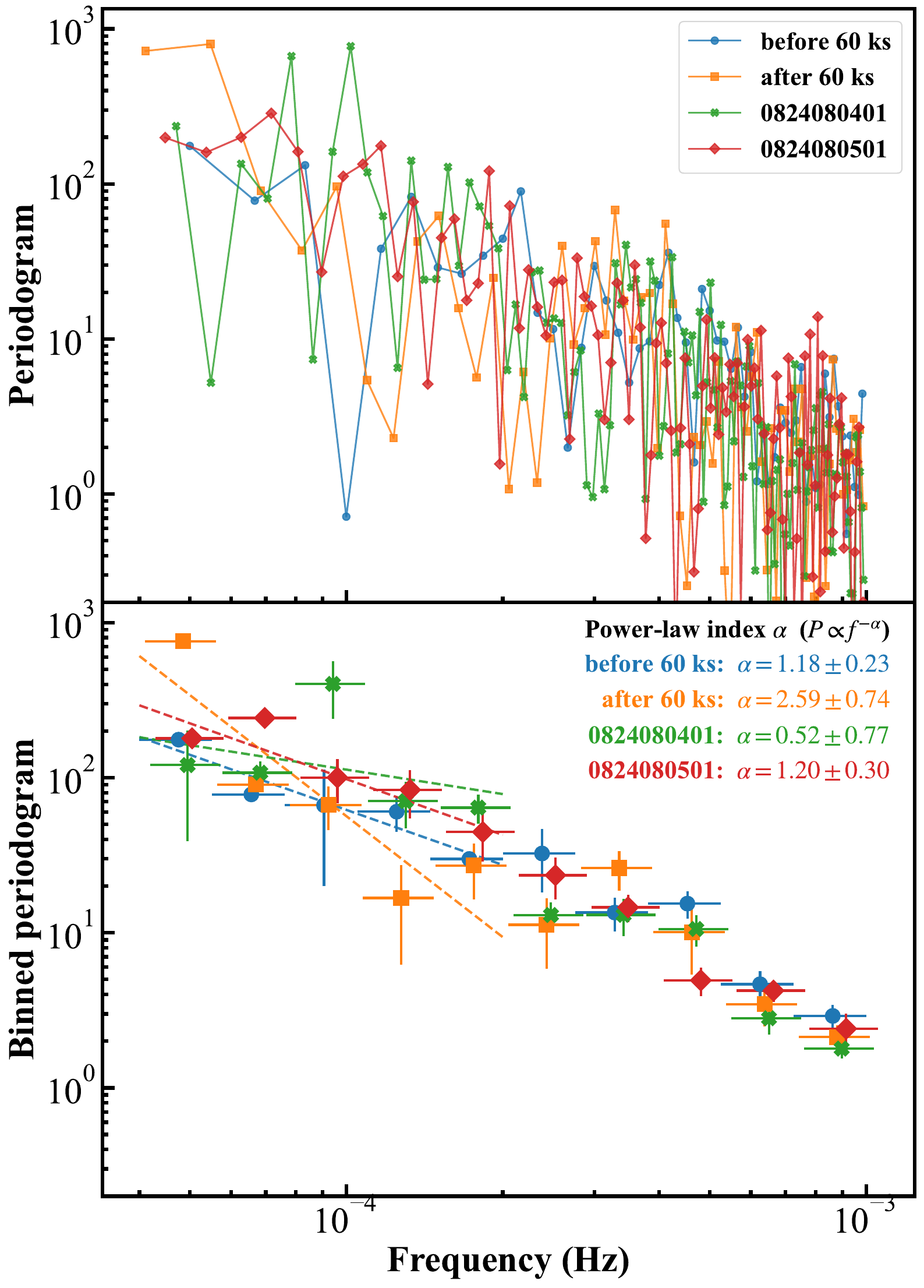}}
\caption{\label{fig:PSD}Power spectral density (PSD) of the 0.5–10 keV light curve for the four observation intervals, color-coded as in Fig. \ref{fig:lags}. The upper panel shows the unbinned periodograms, and the lower panel shows the same data rebinned with finer frequency bins to more clearly illustrate the differences in PSD shape between the intervals. The dashed lines show the best-fit power-law models $P(f) \propto f^{-\alpha}$ fitted to the low-frequency ($f \lesssim 2\times10^{-4}$ Hz) bins of each interval, with the best-fit indices listed in the legend.
}
\end{figure}

\begin{figure}
\centering
\subfloat{\includegraphics[width=0.45\textwidth]{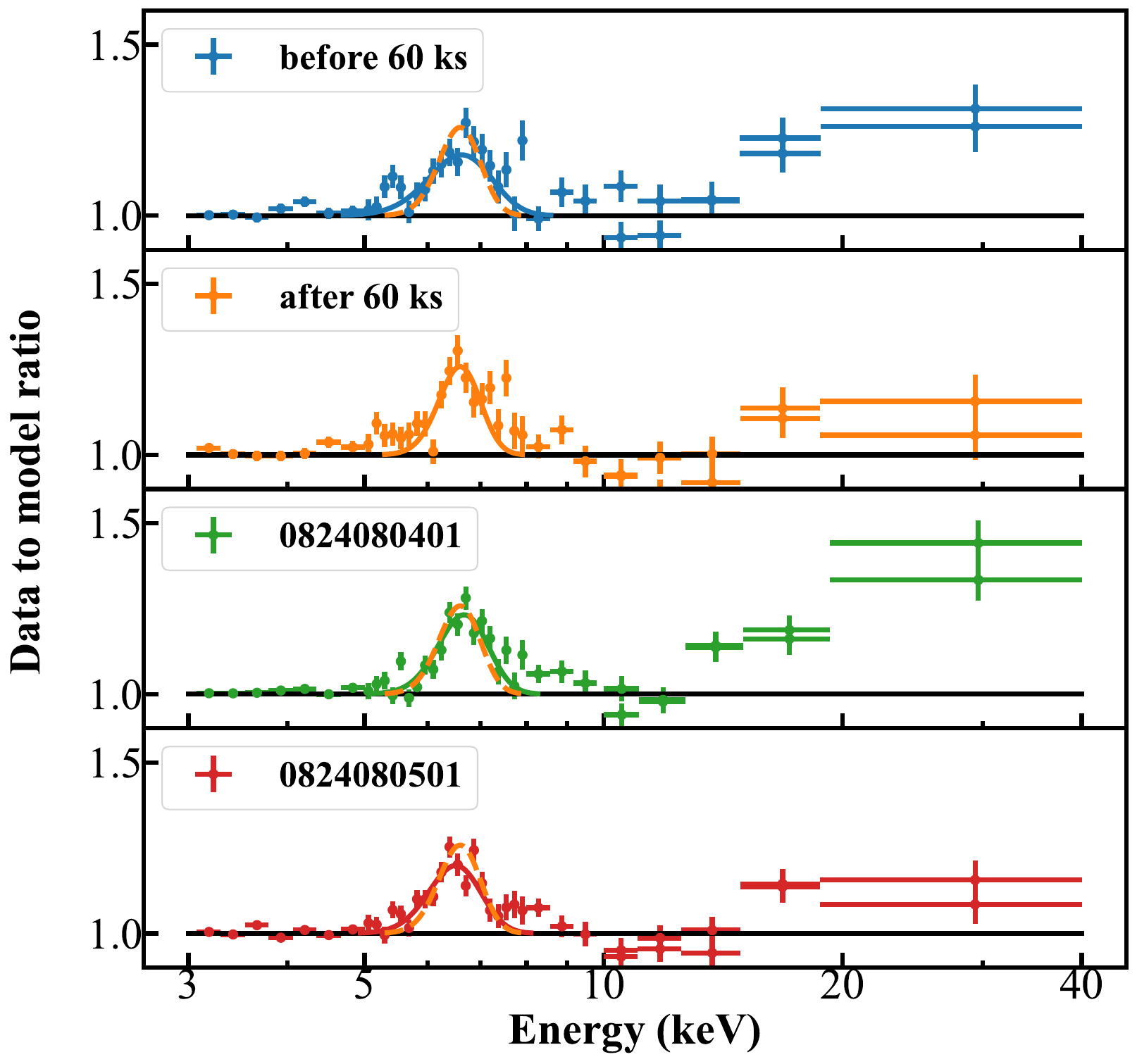}}
\caption{\label{fig:Feline}Data-to-model ratios of the spectra at different epochs. The model is obtained by jointly fitting the XMM–Newton and NuSTAR spectra in the 3–5 and 8–10~keV bands with a model of the form $constant \times powerlaw$, where the constant accounts for cross-calibration between the instruments. For clarity, the data are further rebinned, and the NuSTAR FPMA and FPMB spectra below 10 keV are omitted. The colored lines near 6.4~keV indicate the best-fitting Gaussian components representing the Fe~K$\alpha$ line. The best-fit Gaussian component derived for the lag-detected interval (shown in orange) is overplotted in the other panels for comparison.
}
\end{figure}

The sudden emergence of the hard X-ray lag generally indicates a rapid transformation of the X-ray corona from a compact to a more extended configuration. This interpretation is further supported by two independent observational signatures. First, the light curves (Fig.~\ref{fig:LC}) show that the variability during the lag-detected period appears noticeably smoother than at other epochs, suggesting emission from a larger region. This impression is confirmed by the low-frequency power spectral density (PSD; Fig.~\ref{fig:PSD}), which becomes redder during the same period. To quantify this, we fit a power law $P(f) \propto f^{-\alpha}$ to the low-frequency ($f \lesssim 2\times10^{-4}$ Hz) binned PSD of each interval, finding $\alpha = 1.18\pm0.23$ and $2.59\pm0.74$ for the before- and after-60-ks segments, respectively, confirming that the PSD is steeper during the lag-detected period. Meanwhile, at higher frequencies, the PSDs of all four intervals are consistent with one another.
Second, the profile of the Fe~K$\alpha$ emission line (Fig.~\ref{fig:Feline}) also points to a more extended corona. For each epoch, we fit the 3–5 and 8–10~keV continua with a power law and plot the data-to-model ratio over the Fe~K$\alpha$ and Compton hump regions. Although the spectral quality at high energies is limited due to the steep photon index ($\Gamma \sim 2.5$), the lag-detected period clearly exhibits a narrower Fe line and a weaker Compton hump compared with the pre-transition interval, both consistent with a more extended corona. Meanwhile, the subsequent two \xmm~observations exhibiting weaker hard X-ray lags consistently show a stronger Compton hump and a marginally broader Fe~K$\alpha$ line, suggesting that the corona had contracted again after the lag-detected epoch, though likely less compact than in the initial state.

\par We further attempt to fit the spectra with reflection models to directly measure the reflection strength. Spectral fitting is performed using \textsc{xspec} with $\chi^{2}$ statistics, in the 0.5--10 keV band for EPIC-pn spectra and 3--40 keV for NuSTAR spectra. Uncertainties on each parameter of interest are estimated using $\Delta\chi^{2} = 2.71$, corresponding to the 90\% confidence level. We jointly fitted the EPIC-pn and NuSTAR spectra of the four periods with \textit{constant$\times$TBabs$\times$zxipcf$\times$(nthComp + relxill)}, where \textit{nthComp} \citep{Zycki_1999} models a Comptonised continuum for the potential soft excess and \textit{relxill} \citep{Dauser_2010,Garcia_2014} models a power-law continuum plus its relativistic reflection component. For each period, the spectral parameters of the EPIC-pn and two NuSTAR (FPMA and FPMB) spectra are linked, with a free multiplicative constant to account for cross-normalisation between instruments. The photon index $\Gamma$, reflection fraction, and normalisations are free to vary between periods, while other parameters, including the disc emissivity, inclination, iron abundance, and ionisation, are tied across all periods. The fit is poor ($\chi^{2}/\mathrm{dof} = 8140/6877$), with prominent residuals above 10 keV, indicating that the reflection component is not adequately modelled. This is likely because (1) the spectrum is very soft ($\Gamma \sim 2.5$), rendering the NuSTAR data above 10 keV statistically limited; and (2) there are complex absorption and emission features below 2 keV and around the Fe~K$\alpha$ region associated with ultra-fast outflows \citep{Xu_2023}.

\par We therefore adopt a different model, \textit{constant$\times$TBabs$\times$(pexrav + zgaussian)}, where \textit{pexrav} \citep{Zdziarski_1996} models a power-law continuum plus its reflection component, with the Fe~K$\alpha$ emission line decoupled from the Compton hump and modelled separately using \textit{zgaussian}. Data below 2 keV are excluded to avoid the complex absorption features associated with the ultra-fast outflow \citep{Xu_2023}. The parameters free to vary between periods are $\Gamma$, the reflection 
strength $R$, the line width $\sigma_{\rm line}$, and the normalisations. The fitting is better ($\chi^{2}/\mathrm{dof} = 5987/5664$), and the results are provided in Table \ref{tab:fitting}. Consistent with Figs.~\ref{fig:spectral_ratio} and \ref{fig:Feline}, the lag-detected period shows a marginally narrower Fe line and a weaker reflection component compared with most other intervals, supporting a spatially more extended corona, albeit not at a statistically significant level. Indeed, all three parameters are consistent with being constant across all periods at the 90\% confidence level.

\begin{deluxetable}{cccc}
\tablecaption{Best-fit spectral parameters from the reflection model.\label{tab:fitting}}
\tablehead{
\colhead{Period} & \colhead{$\Gamma$} & \colhead{$R$} & 
\colhead{$\sigma_{\rm line}$ (keV)}}
\startdata
Before 60 ks & $2.55^{+0.04}_{-0.04}$ & $2.61^{+0.49}_{-0.56}$ & $1.43^{+0.22}_{-0.24}$ \\
After 60 ks  & $2.54^{+0.04}_{-0.04}$ & $2.11^{+0.46}_{-0.54}$ & $1.25^{+0.18}_{-0.20}$ \\
0824080401   & $2.50^{+0.03}_{-0.03}$ & $2.66^{+0.41}_{-0.46}$ & $1.55^{+0.21}_{-0.22}$ \\
0824080501   & $2.45^{+0.03}_{-0.03}$ & $2.15^{+0.38}_{-0.43}$ & $1.46^{+0.20}_{-0.21}$ \\
\enddata
\tablecomments{Best-fit results of the model 
\textit{constant$\times$TBabs$\times$(pexrav + zgaussian)}. 
Spectral fitting is performed over 2--10 keV for EPIC-pn 
spectra and 3--40 keV for NuSTAR spectra.}
\end{deluxetable}

However, we do not observe substantial changes in either the X-ray flux or the power-law spectral index across the transition, which is somehow puzzling. In the following, we further examine whether specific coronal geometries can reproduce all these observed signatures.

\par A widely discussed configuration for the X-ray corona is the lamppost model \citep[e.g.,][]{Martocchia_1996, Dauser_2013}, in which a compact, point-like source located along the black hole spin axis illuminates the accretion disk, producing both the reflected emission and the reverberation lags \citep{Emmanoulopoulos_2014}. Several studies have suggested that an extended lamppost geometry, often represented by two vertically separated point-like coronae, can more naturally reproduce the observed high-frequency reverberation lags and their variability \citep{Chainakun_2017, Nakhonthong_2024}. Within this framework, the sudden emergence of the hard X-ray lag in Mrk~1044 could in principle arise from an increased separation between the two coronal zones (e.g., if the harder, inner region were ejected upward), causing a longer propagation time for upward-moving fluctuations \citep{Schnittman_2013}.

However, relativistic reflection models predict that both the overall flux and the spectral shape should vary substantially with the height of the lamppost corona \citep[e.g., \textsc{relxilllp};][]{Garcia_2014}. This expectation is also consistent with the empirically observed height–luminosity correlations in some sources \citep{Alston_2020}. In contrast, Mrk~1044 shows nearly constant time-averaged flux and spectral shape across the transition, inconsistent with the (two-zone) lamppost scenario. Moreover, no significant high-frequency reverberation soft lag is detected in any of the observations.

\par Meanwhile, a spherical corona can also give rise to low-frequency hard lags and high-frequency soft lags \citep{Chainakun_2019}. In such a geometry, the hard X-ray lag primarily arises from the inward propagation of fluctuations \citep[e.g.,][]{Wilkins_2016} and the additional time required for multiple Compton scatterings to upscatter seed photons to higher energies \citep[in case of high optical depth and large corona size, ][]{Chainakun_2019}. In the simplest case of a homogeneous spherical corona, a longer hard X-ray lag could result either from an expanded corona (and hence a lower optical depth, $\tau$) or from a substantially higher optical depth that increases the average scattering time \citep{Chainakun_2019}.

Because the photon index of the emerging power-law continuum depends mainly on the coronal temperature and $\tau$ \citep{Zdziarski_1996, Zycki_1999}, one would expect a noticeable change in spectral slope when the hard X-ray lag appears—unless the temperature of the corona is simultaneously fine-tuned to compensate. Observationally, the coronal size, temperature, and optical depth often vary in concert with the X-ray flux \citep[e.g.,][]{Wilkins_2015, Zhangjx2018, Kang_2021, Li_2024}. However, the X-ray flux of Mrk~1044 remains nearly constant across the appearance of the hard lag. Therefore, the spherical-corona scenario also appears unable to fully account for the observed behavior in Mrk~1044.

Here we find the “patchy” coronal geometry to be the most plausible explanation, as it can naturally produce the observed hard X-ray lag while keeping the overall flux and spectral shape nearly unchanged. In this scenario, the corona consists of multiple spatially separated emitting regions, such as nanoflares or magnetic loops \citep{Galeev_1979, Haardt_1994, Stern_1995}. The observed variability could arise if the spatial extent of the entire coronal region—where these flares occur—changes with time, thereby modifying the low-frequency lag, while the total power, average temperature, and optical depth of individual flares remain roughly constant. We emphasize, however, that this interpretation is highly speculative. Even if the corona is indeed patchy, the physical mechanism governing the spatial extent of the coronal region remains unclear.

\section{Conclusions}

We report the discovery of a sudden and pronounced emergence of a low-frequency hard X-ray lag ($\sim$ 1.5 ks between 2–10 and 0.5–2 keV) in the narrow-line Seyfert 1 galaxy Mrk~1044, even though its time-averaged flux and spectral shape remain nearly unchanged. The lag is undetectable during the first $\sim$60~ks of the initial XMM–Newton observation, but then emerges abruptly within $\lesssim$10~ks, becoming prominent during the latter half of the exposure and subsequently weakening in the following two observations.

The transition was first identified through a prominent clockwise loop in the time-resolved flux–softness diagram, demonstrating that this method provides an effective and intuitive means to detect hard X-ray lags and their potential rapid evolution.

The sudden appearance of the hard X-ray lag itself points to a rapid expansion of the corona, from a compact to a more extended configuration. Consistently, spectral and timing analyses show that the lag-detected period is characterized by smoother variability, a redder PSD, and a narrower Fe~K$\alpha$ line with a weaker Compton hump, all additional signatures of a temporarily more extended corona.

Neither the (two-zone) lamppost nor the spherical coronal model appears sufficient to explain the emergence of a strong hard lag without concurrent flux or spectral variations. We speculate that a patchy, dynamically evolving corona, composed of spatially separated emitting regions, could be a more plausible explanation.

This finding demonstrates that the X-ray corona in AGNs can undergo rapid geometrical reconfigurations even when its time-averaged emission remains remarkably stable.
Rapid variations in low-frequency hard X-ray lags thus provide a new means to probe the dynamic behavior of AGN coronae, complementary to other diagnostics such as changes in the soft X-ray reverberation lags or in the broad Fe~K$\alpha$ line profile.

 \acknowledgments
 {This work is supported by the National Natural Science Foundation of China (grant Nos. 123B2042, 12533006 \& 12192221), and Guizhou Provincial Major Scientific and Technological Program XKBF (2025)010 and XKBF (2025)011. The research utilized observations obtained with \xmm~, an ESA science mission supported by contributions from ESA Member States and NASA. This research has made use of the \Nu~Data Analysis Software (NuSTARDAS) jointly developed by the ASI Science Data Center (ASDC, Italy) and the California Institute of Technology (USA).}
 
\software{ 
\textsc{pylag}\footnote{\url{https://github.com/wilkinsdr/pyLag}} \citep{Wilkins_2019},
\textsc{xspec} \citep{Arnaud_1996},
\textsc{astropy} \citep{Astropy_2013}, 
\textsc{heasoft} \citep{HEAsoft_2014},
\textsc{GNU parallel} \citep{Tange2011a}}

\bibliography{sample63}{}

\begin{thebibliography}{}
\expandafter\ifx\csname natexlab\endcsname\relax\def\natexlab#1{#1}\fi
\providecommand{\url}[1]{\href{#1}{#1}}
\providecommand{\dodoi}[1]{doi:~\href{http://doi.org/#1}{\nolinkurl{#1}}}
\providecommand{\doeprint}[1]{\href{http://ascl.net/#1}{\nolinkurl{http://ascl.net/#1}}}
\providecommand{\doarXiv}[1]{\href{https://arxiv.org/abs/#1}{\nolinkurl{https://arxiv.org/abs/#1}}}

\bibitem[{{Alston} {et~al.}(2014){Alston}, {Done}, \& {Vaughan}}]{Alston_2014}
{Alston}, W.~N., {Done}, C., \& {Vaughan}, S. 2014, \mnras, 439, 1548,
  \dodoi{10.1093/mnras/stu005}

\bibitem[{{Alston} {et~al.}(2013){Alston}, {Vaughan}, \&
  {Uttley}}]{Alston_2013}
{Alston}, W.~N., {Vaughan}, S., \& {Uttley}, P. 2013, \mnras, 435, 1511,
  \dodoi{10.1093/mnras/stt1391}

\bibitem[{{Alston} {et~al.}(2020){Alston}, {Fabian}, {Kara}, {Parker},
  {Dovciak}, {Pinto}, {Jiang}, {Middleton}, {Miniutti}, {Walton}, {Wilkins},
  {Buisson}, {Caballero-Garcia}, {Cackett}, {De Marco}, {Gallo}, {Lohfink},
  {Reynolds}, {Uttley}, {Young}, \& {Zogbhi}}]{Alston_2020}
{Alston}, W.~N., {Fabian}, A.~C., {Kara}, E., {et~al.} 2020, Nature Astronomy,
  2, \dodoi{10.1038/s41550-019-1002-x}

\bibitem[{{Ar{\'e}valo} {et~al.}(2008){Ar{\'e}valo}, {McHardy}, \&
  {Summons}}]{Arevalo_2008}
{Ar{\'e}valo}, P., {McHardy}, I.~M., \& {Summons}, D.~P. 2008, \mnras, 388,
  211, \dodoi{10.1111/j.1365-2966.2008.13367.x}

\bibitem[{{Ar{\'e}valo} \& {Uttley}(2006)}]{Arevalo_2006}
{Ar{\'e}valo}, P., \& {Uttley}, P. 2006, \mnras, 367, 801,
  \dodoi{10.1111/j.1365-2966.2006.09989.x}

\bibitem[{{Arnaud}(1996)}]{Arnaud_1996}
{Arnaud}, K.~A. 1996, in Astronomical Society of the Pacific Conference Series,
  Vol. 101, Astronomical Data Analysis Software and Systems V, ed. G.~H.
  {Jacoby} \& J.~{Barnes}, 17

\bibitem[{{Astropy Collaboration} {et~al.}(2013){Astropy Collaboration},
  {Robitaille}, {Tollerud}, {Greenfield}, {Droettboom}, {Bray}, {Aldcroft},
  {Davis}, {Ginsburg}, {Price-Whelan}, {Kerzendorf}, {Conley}, {Crighton},
  {Barbary}, {Muna}, {Ferguson}, {Grollier}, {Parikh}, {Nair}, {Unther},
  {Deil}, {Woillez}, {Conseil}, {Kramer}, {Turner}, {Singer}, {Fox}, {Weaver},
  {Zabalza}, {Edwards}, {Azalee Bostroem}, {Burke}, {Casey}, {Crawford},
  {Dencheva}, {Ely}, {Jenness}, {Labrie}, {Lim}, {Pierfederici}, {Pontzen},
  {Ptak}, {Refsdal}, {Servillat}, \& {Streicher}}]{Astropy_2013}
{Astropy Collaboration}, {Robitaille}, T.~P., {Tollerud}, E.~J., {et~al.} 2013,
  \aap, 558, A33, \dodoi{10.1051/0004-6361/201322068}

\bibitem[{{Barua} {et~al.}(2023){Barua}, {Adegoke}, {Misra}, {Pawar},
  {Jithesh}, \& {Medhi}}]{Barua_2023}
{Barua}, S., {Adegoke}, O.~K., {Misra}, R., {et~al.} 2023, \apj, 958, 46,
  \dodoi{10.3847/1538-4357/acf464}

\bibitem[{{Bianchi} {et~al.}(2009){Bianchi}, {Guainazzi}, {Matt}, {Fonseca
  Bonilla}, \& {Ponti}}]{Bianchi_2009}
{Bianchi}, S., {Guainazzi}, M., {Matt}, G., {Fonseca Bonilla}, N., \& {Ponti},
  G. 2009, \aap, 495, 421, \dodoi{10.1051/0004-6361:200810620}

\bibitem[{{Boissay} {et~al.}(2016){Boissay}, {Ricci}, \&
  {Paltani}}]{Boissay_2016}
{Boissay}, R., {Ricci}, C., \& {Paltani}, S. 2016, \aap, 588, A70,
  \dodoi{10.1051/0004-6361/201526982}

\bibitem[{{Caballero-Garc{\'\i}a} {et~al.}(2018){Caballero-Garc{\'\i}a},
  {Papadakis}, {Dov{\v{c}}iak}, {Bursa}, {Epitropakis}, {Karas}, \&
  {Svoboda}}]{Caballero_2018}
{Caballero-Garc{\'\i}a}, M.~D., {Papadakis}, I.~E., {Dov{\v{c}}iak}, M.,
  {et~al.} 2018, \mnras, 480, 2650, \dodoi{10.1093/mnras/sty1990}

\bibitem[{{Chainakun} {et~al.}(2019){Chainakun}, {Watcharangkool}, {Young}, \&
  {Hancock}}]{Chainakun_2019}
{Chainakun}, P., {Watcharangkool}, A., {Young}, A.~J., \& {Hancock}, S. 2019,
  \mnras, 487, 667, \dodoi{10.1093/mnras/stz1319}

\bibitem[{{Chainakun} \& {Young}(2017)}]{Chainakun_2017}
{Chainakun}, P., \& {Young}, A.~J. 2017, \mnras, 465, 3965,
  \dodoi{10.1093/mnras/stw2964}

\bibitem[{{Chen} {et~al.}(2025{\natexlab{a}}){Chen}, {Wang}, {Kang}, {Kang},
  {Sou}, {Liu}, {Cai}, \& {Su}}]{Chen_2025}
{Chen}, S.-J., {Wang}, J.-X., {Kang}, J.-L., {et~al.} 2025{\natexlab{a}}, \apj,
  980, 23, \dodoi{10.3847/1538-4357/ada035}

\bibitem[{{Chen} {et~al.}(2025{\natexlab{b}}){Chen}, {Wang}, {Kang}, {Kang},
  {Sou}, {Liu}, {Cai}, \& {Su}}]{Chen_2025b}
---. 2025{\natexlab{b}}, arXiv e-prints, arXiv:2510.16525,
  \dodoi{10.48550/arXiv.2510.16525}

\bibitem[{{Crummy} {et~al.}(2006){Crummy}, {Fabian}, {Gallo}, \&
  {Ross}}]{Crummy_2006}
{Crummy}, J., {Fabian}, A.~C., {Gallo}, L., \& {Ross}, R.~R. 2006, \mnras, 365,
  1067, \dodoi{10.1111/j.1365-2966.2005.09844.x}

\bibitem[{{Czerny} {et~al.}(2003){Czerny}, {Niko{\l}ajuk},
  {R{\'o}{\.z}a{\'n}ska}, {Dumont}, {Loska}, \& {Zycki}}]{Czerny_2003}
{Czerny}, B., {Niko{\l}ajuk}, M., {R{\'o}{\.z}a{\'n}ska}, A., {et~al.} 2003,
  \aap, 412, 317, \dodoi{10.1051/0004-6361:20031441}

\bibitem[{{Dauser} {et~al.}(2013){Dauser}, {Garcia}, {Wilms}, {B{\"o}ck},
  {Brenneman}, {Falanga}, {Fukumura}, \& {Reynolds}}]{Dauser_2013}
{Dauser}, T., {Garcia}, J., {Wilms}, J., {et~al.} 2013, \mnras, 430, 1694,
  \dodoi{10.1093/mnras/sts710}

\bibitem[{{Dauser} {et~al.}(2010){Dauser}, {Wilms}, {Reynolds}, \&
  {Brenneman}}]{Dauser_2010}
{Dauser}, T., {Wilms}, J., {Reynolds}, C.~S., \& {Brenneman}, L.~W. 2010,
  \mnras, 409, 1534, \dodoi{10.1111/j.1365-2966.2010.17393.x}

\bibitem[{{De Marco} {et~al.}(2013){De Marco}, {Ponti}, {Cappi}, {Dadina},
  {Uttley}, {Cackett}, {Fabian}, \& {Miniutti}}]{DeMarco_2013}
{De Marco}, B., {Ponti}, G., {Cappi}, M., {et~al.} 2013, \mnras, 431, 2441,
  \dodoi{10.1093/mnras/stt339}

\bibitem[{{Dewangan} {et~al.}(2007){Dewangan}, {Griffiths}, {Dasgupta}, \&
  {Rao}}]{Dewangan_2007}
{Dewangan}, G.~C., {Griffiths}, R.~E., {Dasgupta}, S., \& {Rao}, A.~R. 2007,
  \apj, 671, 1284, \dodoi{10.1086/523683}

\bibitem[{{Done} {et~al.}(2012){Done}, {Davis}, {Jin}, {Blaes}, \&
  {Ward}}]{Done_2012}
{Done}, C., {Davis}, S.~W., {Jin}, C., {Blaes}, O., \& {Ward}, M. 2012, \mnras,
  420, 1848, \dodoi{10.1111/j.1365-2966.2011.19779.x}

\bibitem[{{Du} {et~al.}(2015){Du}, {Hu}, {Lu}, {Huang}, {Cheng}, {Qiu}, {Li},
  {Zhang}, {Fan}, {Bai}, {Bian}, {Yuan}, {Kaspi}, {Ho}, {Netzer}, {Wang}, \&
  {SEAMBH Collaboration}}]{Du_2015}
{Du}, P., {Hu}, C., {Lu}, K.-X., {et~al.} 2015, \apj, 806, 22,
  \dodoi{10.1088/0004-637X/806/1/22}

\bibitem[{{Emmanoulopoulos} {et~al.}(2011){Emmanoulopoulos}, {McHardy}, \&
  {Papadakis}}]{Emmanoulopoulos_2011}
{Emmanoulopoulos}, D., {McHardy}, I.~M., \& {Papadakis}, I.~E. 2011, \mnras,
  416, L94, \dodoi{10.1111/j.1745-3933.2011.01106.x}

\bibitem[{{Emmanoulopoulos} {et~al.}(2014){Emmanoulopoulos}, {Papadakis},
  {Dov{\v{c}}iak}, \& {McHardy}}]{Emmanoulopoulos_2014}
{Emmanoulopoulos}, D., {Papadakis}, I.~E., {Dov{\v{c}}iak}, M., \& {McHardy},
  I.~M. 2014, \mnras, 439, 3931, \dodoi{10.1093/mnras/stu249}

\bibitem[{{Epitropakis} \& {Papadakis}(2016)}]{Epitropakis_2016}
{Epitropakis}, A., \& {Papadakis}, I.~E. 2016, \aap, 591, A113,
  \dodoi{10.1051/0004-6361/201527665}

\bibitem[{{Epitropakis} \& {Papadakis}(2017)}]{Epitropakis_2017}
---. 2017, \mnras, 468, 3568, \dodoi{10.1093/mnras/stx612}

\bibitem[{{Fabian} {et~al.}(2009){Fabian}, {Zoghbi}, {Ross}, {Uttley}, {Gallo},
  {Brandt}, {Blustin}, {Boller}, {Caballero-Garcia}, {Larsson}, {Miller},
  {Miniutti}, {Ponti}, {Reis}, {Reynolds}, {Tanaka}, \& {Young}}]{Fabian_2009}
{Fabian}, A.~C., {Zoghbi}, A., {Ross}, R.~R., {et~al.} 2009, \nat, 459, 540,
  \dodoi{10.1038/nature08007}

\bibitem[{{Galeev} {et~al.}(1979){Galeev}, {Rosner}, \& {Vaiana}}]{Galeev_1979}
{Galeev}, A.~A., {Rosner}, R., \& {Vaiana}, G.~S. 1979, \apj, 229, 318,
  \dodoi{10.1086/156957}

\bibitem[{{Gallo} {et~al.}(2021){Gallo}, {Gonzalez}, \& {Miller}}]{Gallo_2021}
{Gallo}, L.~C., {Gonzalez}, A.~G., \& {Miller}, J.~M. 2021, \apjl, 908, L33,
  \dodoi{10.3847/2041-8213/abdcb5}

\bibitem[{{Garc{\'\i}a} {et~al.}(2014){Garc{\'\i}a}, {Dauser}, {Lohfink},
  {Kallman}, {Steiner}, {McClintock}, {Brenneman}, {Wilms}, {Eikmann},
  {Reynolds}, \& {Tombesi}}]{Garcia_2014}
{Garc{\'\i}a}, J., {Dauser}, T., {Lohfink}, A., {et~al.} 2014, \apj, 782, 76,
  \dodoi{10.1088/0004-637X/782/2/76}

\bibitem[{{Garc{\'\i}a} {et~al.}(2019){Garc{\'\i}a}, {Kara}, {Walton},
  {Beuchert}, {Dauser}, {Gatuzz}, {Balokovic}, {Steiner}, {Tombesi}, {Connors},
  {Kallman}, {Harrison}, {Fabian}, {Wilms}, {Stern}, {Lanz}, {Ricci}, \&
  {Ballantyne}}]{Garcia_2019}
{Garc{\'\i}a}, J.~A., {Kara}, E., {Walton}, D., {et~al.} 2019, \apj, 871, 88,
  \dodoi{10.3847/1538-4357/aaf739}

\bibitem[{{George} \& {Fabian}(1991)}]{George_1991}
{George}, I.~M., \& {Fabian}, A.~C. 1991, \mnras, 249, 352,
  \dodoi{10.1093/mnras/249.2.352}

\bibitem[{{Gierli{\'n}ski} \& {Done}(2004)}]{Gierlinski_2004}
{Gierli{\'n}ski}, M., \& {Done}, C. 2004, \mnras, 349, L7,
  \dodoi{10.1111/j.1365-2966.2004.07687.x}

\bibitem[{{Haardt} \& {Maraschi}(1991)}]{Haardt_1991}
{Haardt}, F., \& {Maraschi}, L. 1991, \apjl, 380, L51, \dodoi{10.1086/186171}

\bibitem[{{Haardt} \& {Maraschi}(1993)}]{Haardt_1993}
---. 1993, \apj, 413, 507, \dodoi{10.1086/173020}

\bibitem[{{Haardt} {et~al.}(1994){Haardt}, {Maraschi}, \&
  {Ghisellini}}]{Haardt_1994}
{Haardt}, F., {Maraschi}, L., \& {Ghisellini}, G. 1994, \apjl, 432, L95,
  \dodoi{10.1086/187520}

\bibitem[{{Hancock} {et~al.}(2022){Hancock}, {Young}, \&
  {Chainakun}}]{Hancock_2022}
{Hancock}, S., {Young}, A.~J., \& {Chainakun}, P. 2022, \mnras, 514, 5403,
  \dodoi{10.1093/mnras/stac1653}

\bibitem[{Harrison {et~al.}(2013)Harrison, Craig, Christensen, Hailey, Zhang,
  Boggs, Stern, Cook, Forster, Giommi, Grefenstette, Kim, Kitaguchi, Koglin,
  Madsen, Mao, Miyasaka, Mori, Perri, Pivovaroff, Puccetti, Rana, Westergaard,
  Willis, Zoglauer, An, Bachetti, Barri{\`{e}}re, Bellm, Bhalerao, Brejnholt,
  Fuerst, Liebe, Markwardt, Nynka, Vogel, Walton, Wik, Alexander, Cominsky,
  Hornschemeier, Hornstrup, Kaspi, Madejski, Matt, Molendi, Smith, Tomsick,
  Ajello, Ballantyne, Balokovi{\'{c}}, Barret, Bauer, Blandford, Brandt,
  Brenneman, Chiang, Chakrabarty, Chenevez, Comastri, Dufour, Elvis, Fabian,
  Farrah, Fryer, Gotthelf, Grindlay, Helfand, Krivonos, Meier, Miller,
  Natalucci, Ogle, Ofek, Ptak, Reynolds, Rigby, Tagliaferri, Thorsett,
  Treister, \& Urry}]{Harrison_2013}
Harrison, F.~A., Craig, W.~W., Christensen, F.~E., {et~al.} 2013, The
  Astrophysical Journal, 770, 103, \dodoi{10.1088/0004-637x/770/2/103}

\bibitem[{{Jansen} {et~al.}(2001){Jansen}, {Lumb}, {Altieri}, {Clavel}, {Ehle},
  {Erd}, {Gabriel}, {Guainazzi}, {Gondoin}, {Much}, {Munoz}, {Santos},
  {Schartel}, {Texier}, \& {Vacanti}}]{Jansen_2001}
{Jansen}, F., {Lumb}, D., {Altieri}, B., {et~al.} 2001, \aap, 365, L1,
  \dodoi{10.1051/0004-6361:20000036}

\bibitem[{{Jiang} {et~al.}(2019){Jiang}, {Fabian}, {Dauser}, {Gallo},
  {Garc{\'\i}a}, {Kara}, {Parker}, {Tomsick}, {Walton}, \&
  {Reynolds}}]{Jiang_2019}
{Jiang}, J., {Fabian}, A.~C., {Dauser}, T., {et~al.} 2019, \mnras, 489, 3436,
  \dodoi{10.1093/mnras/stz2326}

\bibitem[{Kang \& Wang(2024)}]{Kang_2024}
Kang, J., \& Wang, J. 2024, JUSTC, 54, 0702, \dodoi{10.52396/JUSTC-2023-0160}

\bibitem[{{Kang} \& {Wang}(2022)}]{Kang_2022}
{Kang}, J.-L., \& {Wang}, J.-X. 2022, \apj, 929, 141,
  \dodoi{10.3847/1538-4357/ac5d49}

\bibitem[{{Kang} {et~al.}(2023){Kang}, {Wang}, \& {Fu}}]{Kang_2023}
{Kang}, J.-L., {Wang}, J.-X., \& {Fu}, S.-Q. 2023, \mnras, 525, 1941,
  \dodoi{10.1093/mnras/stad2364}

\bibitem[{{Kang} {et~al.}(2021){Kang}, {Wang}, \& {Kang}}]{Kang_2021}
{Kang}, J.-L., {Wang}, J.-X., \& {Kang}, W.-Y. 2021, \mnras, 502, 80,
  \dodoi{10.1093/mnras/stab039}

\bibitem[{{Kara} {et~al.}(2016){Kara}, {Alston}, {Fabian}, {Cackett}, {Uttley},
  {Reynolds}, \& {Zoghbi}}]{Kara_2016}
{Kara}, E., {Alston}, W.~N., {Fabian}, A.~C., {et~al.} 2016, \mnras, 462, 511,
  \dodoi{10.1093/mnras/stw1695}

\bibitem[{{Kara} {et~al.}(2013{\natexlab{a}}){Kara}, {Fabian}, {Cackett},
  {Miniutti}, \& {Uttley}}]{Kara_2013b}
{Kara}, E., {Fabian}, A.~C., {Cackett}, E.~M., {Miniutti}, G., \& {Uttley}, P.
  2013{\natexlab{a}}, \mnras, 430, 1408, \dodoi{10.1093/mnras/stt024}

\bibitem[{{Kara} {et~al.}(2013{\natexlab{b}}){Kara}, {Fabian}, {Cackett},
  {Uttley}, {Wilkins}, \& {Zoghbi}}]{Kara_2013}
{Kara}, E., {Fabian}, A.~C., {Cackett}, E.~M., {et~al.} 2013{\natexlab{b}},
  \mnras, 434, 1129, \dodoi{10.1093/mnras/stt1055}

\bibitem[{{Kara} {et~al.}(2015){Kara}, {Zoghbi}, {Marinucci}, {Walton},
  {Fabian}, {Risaliti}, {Boggs}, {Christensen}, {Fuerst}, {Hailey}, {Harrison},
  {Matt}, {Parker}, {Reynolds}, {Stern}, \& {Zhang}}]{Kara_2015}
{Kara}, E., {Zoghbi}, A., {Marinucci}, A., {et~al.} 2015, \mnras, 446, 737,
  \dodoi{10.1093/mnras/stu2136}

\bibitem[{{Koss} {et~al.}(2022){Koss}, {Ricci}, {Trakhtenbrot}, {Oh}, {den
  Brok}, {Mej{\'\i}a-Restrepo}, {Stern}, {Privon}, {Treister}, {Powell},
  {Mushotzky}, {Bauer}, {Ananna}, {Balokovi{\'c}}, {B{\"a}r}, {Becker},
  {Bessiere}, {Burtscher}, {Caglar}, {Congiu}, {Evans}, {Harrison}, {Heida},
  {Ichikawa}, {Kamraj}, {Lamperti}, {Pacucci}, {Ricci}, {Riffel}, {Rojas},
  {Schawinski}, {Temple}, {Urry}, {Veilleux}, \& {Williams}}]{Koss_2022}
{Koss}, M.~J., {Ricci}, C., {Trakhtenbrot}, B., {et~al.} 2022, \apjs, 261, 2,
  \dodoi{10.3847/1538-4365/ac6c05}

\bibitem[{{Kotov} {et~al.}(2001){Kotov}, {Churazov}, \&
  {Gilfanov}}]{Kotov_2001}
{Kotov}, O., {Churazov}, E., \& {Gilfanov}, M. 2001, \mnras, 327, 799,
  \dodoi{10.1046/j.1365-8711.2001.04769.x}

\bibitem[{{Krongold} {et~al.}(2021){Krongold}, {Longinotti}, {Santos-Lle{\'o}},
  {Mathur}, {Peterson}, {Nicastro}, {Gupta}, {Rodr{\'\i}guez-Pascual}, \&
  {El{\'\i}as-Ch{\'a}vez}}]{Krongold_2021}
{Krongold}, Y., {Longinotti}, A.~L., {Santos-Lle{\'o}}, M., {et~al.} 2021,
  \apj, 917, 39, \dodoi{10.3847/1538-4357/ac0977}

\bibitem[{{Li} {et~al.}(2024){Li}, {Ho}, {Ricci}, \& {Trakhtenbrot}}]{Li_2024}
{Li}, R., {Ho}, L.~C., {Ricci}, C., \& {Trakhtenbrot}, B. 2024, \apj, 975, 50,
  \dodoi{10.3847/1538-4357/ad77a5}

\bibitem[{{Liao} {et~al.}(2024){Liao}, {Wang}, {Kang}, {Li}, \&
  {Zhou}}]{Liao_2024}
{Liao}, M., {Wang}, J., {Kang}, J., {Li}, X., \& {Zhou}, M. 2024, \mnras, 528,
  2742, \dodoi{10.1093/mnras/stae122}

\bibitem[{{Lobban} {et~al.}(2018){Lobban}, {Vaughan}, {Pounds}, \&
  {Reeves}}]{Lobban_2018}
{Lobban}, A.~P., {Vaughan}, S., {Pounds}, K., \& {Reeves}, J.~N. 2018, \mnras,
  476, 225, \dodoi{10.1093/mnras/sty123}

\bibitem[{{Mallick} {et~al.}(2021){Mallick}, {Wilkins}, {Alston}, {Markowitz},
  {De Marco}, {Parker}, {Lohfink}, \& {Stalin}}]{Mallick_2021}
{Mallick}, L., {Wilkins}, D.~R., {Alston}, W.~N., {et~al.} 2021, \mnras, 503,
  3775, \dodoi{10.1093/mnras/stab627}

\bibitem[{{Mallick} {et~al.}(2018){Mallick}, {Alston}, {Parker}, {Fabian},
  {Pinto}, {Dewangan}, {Markowitz}, {Gandhi}, {Kembhavi}, \&
  {Misra}}]{Mallick_2018}
{Mallick}, L., {Alston}, W.~N., {Parker}, M.~L., {et~al.} 2018, \mnras, 479,
  615, \dodoi{10.1093/mnras/sty1487}

\bibitem[{{Marinucci} {et~al.}(2014){Marinucci}, {Matt}, {Kara}, {Miniutti},
  {Elvis}, {Arevalo}, {Ballantyne}, {Balokovi{\'c}}, {Bauer}, {Brenneman},
  {Boggs}, {Cappi}, {Christensen}, {Craig}, {Fabian}, {Fuerst}, {Hailey},
  {Harrison}, {Risaliti}, {Reynolds}, {Stern}, {Walton}, \&
  {Zhang}}]{Marinucci_2014}
{Marinucci}, A., {Matt}, G., {Kara}, E., {et~al.} 2014, \mnras, 440, 2347,
  \dodoi{10.1093/mnras/stu404}

\bibitem[{{Markowitz} {et~al.}(2003){Markowitz}, {Edelson}, \&
  {Vaughan}}]{Markowitz_2003}
{Markowitz}, A., {Edelson}, R., \& {Vaughan}, S. 2003, \apj, 598, 935,
  \dodoi{10.1086/379103}

\bibitem[{{Martocchia} \& {Matt}(1996)}]{Martocchia_1996}
{Martocchia}, A., \& {Matt}, G. 1996, \mnras, 282, L53,
  \dodoi{10.1093/mnras/282.4.L53}

\bibitem[{{McHardy} {et~al.}(2004){McHardy}, {Papadakis}, {Uttley}, {Page}, \&
  {Mason}}]{McHardy_2004}
{McHardy}, I.~M., {Papadakis}, I.~E., {Uttley}, P., {Page}, M.~J., \& {Mason},
  K.~O. 2004, \mnras, 348, 783, \dodoi{10.1111/j.1365-2966.2004.07376.x}

\bibitem[{{Nakhonthong} {et~al.}(2024){Nakhonthong}, {Chainakun}, {Luangtip},
  \& {Young}}]{Nakhonthong_2024}
{Nakhonthong}, N., {Chainakun}, P., {Luangtip}, W., \& {Young}, A.~J. 2024,
  \mnras, 530, 1894, \dodoi{10.1093/mnras/stae978}

\bibitem[{{Nandra} \& {Pounds}(1994)}]{Nandra_1994}
{Nandra}, K., \& {Pounds}, K.~A. 1994, \mnras, 268, 405,
  \dodoi{10.1093/mnras/268.2.405}

\bibitem[{{Nasa High Energy Astrophysics Science Archive Research Center
  (Heasarc)}(2014)}]{HEAsoft_2014}
{Nasa High Energy Astrophysics Science Archive Research Center (Heasarc)}.
  2014, {HEAsoft: Unified Release of FTOOLS and XANADU}, Astrophysics Source
  Code Library, record ascl:1408.004.
\newblock \doeprint{1408.004}

\bibitem[{{Panagiotou} \& {Walter}(2020)}]{Panagiotou_2020}
{Panagiotou}, C., \& {Walter}, R. 2020, \aap, 640, A31,
  \dodoi{10.1051/0004-6361/201937390}

\bibitem[{{Papadakis} {et~al.}(2001){Papadakis}, {Nandra}, \&
  {Kazanas}}]{Papadakis_2001}
{Papadakis}, I.~E., {Nandra}, K., \& {Kazanas}, D. 2001, \apjl, 554, L133,
  \dodoi{10.1086/321722}

\bibitem[{{Petrucci} {et~al.}(2020){Petrucci}, {Gronkiewicz}, {Rozanska},
  {Belmont}, {Bianchi}, {Czerny}, {Matt}, {Malzac}, {Middei}, {De Rosa},
  {Ursini}, \& {Cappi}}]{Petrucci_2020}
{Petrucci}, P.~O., {Gronkiewicz}, D., {Rozanska}, A., {et~al.} 2020, \aap, 634,
  A85, \dodoi{10.1051/0004-6361/201937011}

\bibitem[{{Pounds} {et~al.}(1990){Pounds}, {Nandra}, {Stewart}, {George}, \&
  {Fabian}}]{Pounds_1990}
{Pounds}, K.~A., {Nandra}, K., {Stewart}, G.~C., {George}, I.~M., \& {Fabian},
  A.~C. 1990, \nat, 344, 132, \dodoi{10.1038/344132a0}

\bibitem[{Priestley(1981)}]{Priestley_1981}
Priestley, M.~B. 1981, Spectral Analysis and Time Series (London: Academic
  Press)

\bibitem[{{Ren} {et~al.}(2025){Ren}, {Wang}, \& {Kang}}]{Ren_2025}
{Ren}, L.-X., {Wang}, J.-X., \& {Kang}, J.-L. 2025, Research in Astronomy and
  Astrophysics, 25, 015009, \dodoi{10.1088/1674-4527/ad9386}

\bibitem[{{Ricci} {et~al.}(2011){Ricci}, {Walter}, {Courvoisier}, \&
  {Paltani}}]{Ricci_2011}
{Ricci}, C., {Walter}, R., {Courvoisier}, T.~J.~L., \& {Paltani}, S. 2011,
  \aap, 532, A102, \dodoi{10.1051/0004-6361/201016409}

\bibitem[{{Schnittman} {et~al.}(2013){Schnittman}, {Krolik}, \&
  {Noble}}]{Schnittman_2013}
{Schnittman}, J.~D., {Krolik}, J.~H., \& {Noble}, S.~C. 2013, \apj, 769, 156,
  \dodoi{10.1088/0004-637X/769/2/156}

\bibitem[{{Sobolewska} \& {Papadakis}(2009)}]{Sobolewska_2009}
{Sobolewska}, M.~A., \& {Papadakis}, I.~E. 2009, \mnras, 399, 1597,
  \dodoi{10.1111/j.1365-2966.2009.15382.x}

\bibitem[{{Stern} {et~al.}(1995){Stern}, {Poutanen}, {Svensson}, {Sikora}, \&
  {Begelman}}]{Stern_1995}
{Stern}, B.~E., {Poutanen}, J., {Svensson}, R., {Sikora}, M., \& {Begelman},
  M.~C. 1995, \apjl, 449, L13, \dodoi{10.1086/309617}

\bibitem[{{Str{\"u}der} {et~al.}(2001){Str{\"u}der}, {Briel}, {Dennerl},
  {Hartmann}, {Kendziorra}, {Meidinger}, {Pfeffermann}, {Reppin}, {Aschenbach},
  {Bornemann}, {Br{\"a}uninger}, {Burkert}, {Elender}, {Freyberg}, {Haberl},
  {Hartner}, {Heuschmann}, {Hippmann}, {Kastelic}, {Kemmer}, {Kettenring},
  {Kink}, {Krause}, {M{\"u}ller}, {Oppitz}, {Pietsch}, {Popp}, {Predehl},
  {Read}, {Stephan}, {St{\"o}tter}, {Tr{\"u}mper}, {Holl}, {Kemmer}, {Soltau},
  {St{\"o}tter}, {Weber}, {Weichert}, {von Zanthier}, {Carathanassis}, {Lutz},
  {Richter}, {Solc}, {B{\"o}ttcher}, {Kuster}, {Staubert}, {Abbey}, {Holland},
  {Turner}, {Balasini}, {Bignami}, {La Palombara}, {Villa}, {Buttler},
  {Gianini}, {Lain{\'e}}, {Lumb}, \& {Dhez}}]{Struder_2001}
{Str{\"u}der}, L., {Briel}, U., {Dennerl}, K., {et~al.} 2001, \aap, 365, L18,
  \dodoi{10.1051/0004-6361:20000066}

\bibitem[{{Tanaka} {et~al.}(1995){Tanaka}, {Nandra}, {Fabian}, {Inoue},
  {Otani}, {Dotani}, {Hayashida}, {Iwasawa}, {Kii}, {Kunieda}, {Makino}, \&
  {Matsuoka}}]{Tanaka_1995}
{Tanaka}, Y., {Nandra}, K., {Fabian}, A.~C., {et~al.} 1995, \nat, 375, 659,
  \dodoi{10.1038/375659a0}

\bibitem[{Tange(2011)}]{Tange2011a}
Tange, O. 2011, ;login: The USENIX Magazine, 36, 42,
  \dodoi{10.5281/zenodo.16303}

\bibitem[{{Timmer} \& {K{\"o}nig}(1995)}]{Timmer_1995}
{Timmer}, J., \& {K{\"o}nig}, M. 1995, \aap, 300, 707

\bibitem[{{Ursini} {et~al.}(2020){Ursini}, {Dov{\v{c}}iak}, {Zhang}, {Matt},
  {Petrucci}, \& {Done}}]{Ursini_2020b}
{Ursini}, F., {Dov{\v{c}}iak}, M., {Zhang}, W., {et~al.} 2020, \aap, 644, A132,
  \dodoi{10.1051/0004-6361/202039158}

\bibitem[{{Uttley} {et~al.}(2014){Uttley}, {Cackett}, {Fabian}, {Kara}, \&
  {Wilkins}}]{Uttley_2014}
{Uttley}, P., {Cackett}, E.~M., {Fabian}, A.~C., {Kara}, E., \& {Wilkins},
  D.~R. 2014, \aapr, 22, 72, \dodoi{10.1007/s00159-014-0072-0}

\bibitem[{{Walton} {et~al.}(2013){Walton}, {Nardini}, {Fabian}, {Gallo}, \&
  {Reis}}]{Walton_2013}
{Walton}, D.~J., {Nardini}, E., {Fabian}, A.~C., {Gallo}, L.~C., \& {Reis},
  R.~C. 2013, \mnras, 428, 2901, \dodoi{10.1093/mnras/sts227}

\bibitem[{{Walton} {et~al.}(2014){Walton}, {Risaliti}, {Harrison}, {Fabian},
  {Miller}, {Arevalo}, {Ballantyne}, {Boggs}, {Brenneman}, {Christensen},
  {Craig}, {Elvis}, {Fuerst}, {Gandhi}, {Grefenstette}, {Hailey}, {Kara},
  {Luo}, {Madsen}, {Marinucci}, {Matt}, {Parker}, {Reynolds}, {Rivers}, {Ross},
  {Stern}, \& {Zhang}}]{Walton_2014}
{Walton}, D.~J., {Risaliti}, G., {Harrison}, F.~A., {et~al.} 2014, \apj, 788,
  76, \dodoi{10.1088/0004-637X/788/1/76}

\bibitem[{{Wang} {et~al.}(1999){Wang}, {Zhou}, \& {Wang}}]{Wang_1999}
{Wang}, J.-X., {Zhou}, Y.-Y., \& {Wang}, T.-G. 1999, \apjl, 523, L129,
  \dodoi{10.1086/312268}

\bibitem[{{Weng} {et~al.}(2021){Weng}, {Cai}, {Zhang}, {Zhang}, {Chen},
  {Huang}, \& {Tao}}]{Weng_2021}
{Weng}, S.-S., {Cai}, Z.-Y., {Zhang}, S.-N., {et~al.} 2021, \apjl, 915, L15,
  \dodoi{10.3847/2041-8213/ac0a7b}

\bibitem[{{Wilkins}(2019)}]{Wilkins_2019}
{Wilkins}, D.~R. 2019, \mnras, 489, 1957, \dodoi{10.1093/mnras/stz2269}

\bibitem[{{Wilkins}(2023)}]{Wilkins_2023}
---. 2023, \mnras, 526, 3441, \dodoi{10.1093/mnras/stad2936}

\bibitem[{{Wilkins} {et~al.}(2016){Wilkins}, {Cackett}, {Fabian}, \&
  {Reynolds}}]{Wilkins_2016}
{Wilkins}, D.~R., {Cackett}, E.~M., {Fabian}, A.~C., \& {Reynolds}, C.~S. 2016,
  \mnras, 458, 200, \dodoi{10.1093/mnras/stw276}

\bibitem[{{Wilkins} \& {Fabian}(2013)}]{Wilkins_2013}
{Wilkins}, D.~R., \& {Fabian}, A.~C. 2013, \mnras, 430, 247,
  \dodoi{10.1093/mnras/sts591}

\bibitem[{{Wilkins} \& {Gallo}(2015)}]{Wilkins_2015}
{Wilkins}, D.~R., \& {Gallo}, L.~C. 2015, \mnras, 449, 129,
  \dodoi{10.1093/mnras/stv162}

\bibitem[{{Wu} {et~al.}(2020){Wu}, {Wang}, {Cai}, {Kang}, {Liu}, \&
  {Cai}}]{Wu_2020}
{Wu}, Y.-J., {Wang}, J.-X., {Cai}, Z.-Y., {et~al.} 2020, Science China Physics,
  Mechanics, and Astronomy, 63, 129512, \dodoi{10.1007/s11433-020-1611-7}

\bibitem[{{Xu} {et~al.}(2023){Xu}, {Pinto}, {Rogantini}, {Bianchi},
  {Guainazzi}, {Kara}, {Jin}, \& {Cusumano}}]{Xu_2023}
{Xu}, Y., {Pinto}, C., {Rogantini}, D., {et~al.} 2023, \mnras, 523, 2158,
  \dodoi{10.1093/mnras/stad1565}

\bibitem[{{Yu} {et~al.}(2025){Yu}, {Wilkins}, \& {Allen}}]{Yu_2025}
{Yu}, Z., {Wilkins}, D., \& {Allen}, S.~W. 2025, \apj, 989, 212,
  \dodoi{10.3847/1538-4357/adef4f}

\bibitem[{{Zdziarski} {et~al.}(1996){Zdziarski}, {Johnson}, \&
  {Magdziarz}}]{Zdziarski_1996}
{Zdziarski}, A.~A., {Johnson}, W.~N., \& {Magdziarz}, P. 1996, \mnras, 283,
  193, \dodoi{10.1093/mnras/283.1.193}

\bibitem[{{Zdziarski} {et~al.}(2000){Zdziarski}, {Poutanen}, \&
  {Johnson}}]{Zdziarski_2000}
{Zdziarski}, A.~A., {Poutanen}, J., \& {Johnson}, W.~N. 2000, \apj, 542, 703,
  \dodoi{10.1086/317046}

\bibitem[{{Zhang} {et~al.}(2018){Zhang}, {Wang}, \& {Zhu}}]{Zhangjx2018}
{Zhang}, J.-X., {Wang}, J.-X., \& {Zhu}, F.-F. 2018, \apj, 863, 71,
  \dodoi{10.3847/1538-4357/aacf92}

\bibitem[{{Zhou} {et~al.}(2025){Zhou}, {Mao}, {Zhang}, {Patruno}, {Bozzo},
  {Xu}, {Santangelo}, {Zane}, {Zhang}, {Feng}, {Cavecchi}, {de Marco}, {Fan},
  {Hou}, {Jiang}, {Romano}, {Sala}, {Tao}, {Veledina}, {Vink}, {Wang}, {Wang},
  {Wang}, {Weng}, {Wu}, {Xie}, {Zhang}, {Zhang}, {Zhao}, {Zheng}, {Barua},
  {Chen}, {Chen}, {Chen}, {Chen}, {Chen}, {Cheng}, {Chi}, {Cui}, {de Martino},
  {Deng}, {Ducci}, {Farinelli}, {Feng}, {Ge}, {Gu}, {Guo}, {Han}, {Hu},
  {Huang}, {in't Zand}, {Ji}, {Kang}, {Kini}, {Li}, {Li}, {Liu}, {Liu}, {Liu},
  {Lyu}, {Marino}, {Markowitz}, {Mezcua}, {Middleton}, {Mou}, {Ng}, {Papitto},
  {Pei}, {Peng}, {Poutanen}, {Shui}, {Simone}, {Su}, {Tan}, {Wang}, {Wang},
  {Wang}, {Wang}, {Wang}, {Wang}, {Wang}, {Wu}, {Xiao}, {Xiong}, {Xu}, {Xue},
  {Yan}, {Yang}, {Yang}, {Yang}, {Ye}, {Yu}, {Yuan}, {Zhang}, {Zhang}, {Zhao},
  {Zhao}, {Zheng}, {Zheng}, \& {Zuo}}]{Zhou_2025SCPMA}
{Zhou}, P., {Mao}, J., {Zhang}, L., {et~al.} 2025, Science China Physics,
  Mechanics, and Astronomy, 68, 119507, \dodoi{10.1007/s11433-025-2799-0}

\bibitem[{{Zoghbi} {et~al.}(2012){Zoghbi}, {Fabian}, {Reynolds}, \&
  {Cackett}}]{Zoghbi_2012}
{Zoghbi}, A., {Fabian}, A.~C., {Reynolds}, C.~S., \& {Cackett}, E.~M. 2012,
  \mnras, 422, 129, \dodoi{10.1111/j.1365-2966.2012.20587.x}

\bibitem[{{Zoghbi} {et~al.}(2010){Zoghbi}, {Fabian}, {Uttley}, {Miniutti},
  {Gallo}, {Reynolds}, {Miller}, \& {Ponti}}]{Zoghbi_2010}
{Zoghbi}, A., {Fabian}, A.~C., {Uttley}, P., {et~al.} 2010, \mnras, 401, 2419,
  \dodoi{10.1111/j.1365-2966.2009.15816.x}

\bibitem[{{Zoghbi} \& {Miller}(2023)}]{Zoghbi_2023}
{Zoghbi}, A., \& {Miller}, J.~M. 2023, \apj, 957, 69,
  \dodoi{10.3847/1538-4357/acfb85}

\bibitem[{{Zoghbi} {et~al.}(2021){Zoghbi}, {Miller}, \&
  {Cackett}}]{Zoghbi_2021}
{Zoghbi}, A., {Miller}, J.~M., \& {Cackett}, E. 2021, \apj, 912, 42,
  \dodoi{10.3847/1538-4357/abebd9}

\bibitem[{{Zoghbi} {et~al.}(2014){Zoghbi}, {Cackett}, {Reynolds}, {Kara},
  {Harrison}, {Fabian}, {Lohfink}, {Matt}, {Balokovic}, {Boggs}, {Christensen},
  {Craig}, {Hailey}, {Stern}, \& {Zhang}}]{Zoghbi_2014}
{Zoghbi}, A., {Cackett}, E.~M., {Reynolds}, C., {et~al.} 2014, \apj, 789, 56,
  \dodoi{10.1088/0004-637X/789/1/56}

\bibitem[{{{\.Z}ycki} {et~al.}(1999){{\.Z}ycki}, {Done}, \&
  {Smith}}]{Zycki_1999}
{{\.Z}ycki}, P.~T., {Done}, C., \& {Smith}, D.~A. 1999, \mnras, 309, 561,
  \dodoi{10.1046/j.1365-8711.1999.02885.x}

\end{thebibliography}

\end{document}